\documentclass{egpubl-arxiv}

\JournalPaper         
\usepackage[T1]{fontenc}
\usepackage{dfadobe}

\biberVersion
\BibtexOrBiblatex
\usepackage[backend=biber,bibstyle=EG,citestyle=alphabetic,backref=true]{biblatex}
\electronicVersion
\PrintedOrElectronic

\ifpdf \usepackage[pdftex]{graphicx} \pdfcompresslevel=9
\else \usepackage[dvips]{graphicx} \fi

\usepackage{egweblnk}
\title[Path Guiding Using Spatio-Directional Mixture Models]%
      {Path Guiding Using Spatio-Directional Mixture Models}

\author[A. Dodik \& M. Papas \& C. \"Oztireli \& T. M\"uller]
{\parbox{\textwidth}{\centering A. Dodik$^{1}$ \& M. Papas$^{2}$ \& C. \"Oztireli$^{3}$ \& T. M\"uller$^{4}$}
        \\
{\parbox{\textwidth}{\centering $^1$Facebook$^\dagger$ $\qquad$ $^2$Disney Research|Studios $\qquad$ $^3$University of Cambridge $\qquad$ $^4$NVIDIA
       }
}
}

\usepackage{mathtools}
\gdef\cropInsets{0}

\usepackage{animate}

\usepackage{booktabs} 
\usepackage{tabularx}
\usepackage{xargs}
\usepackage{mathtools}
\usepackage{amsmath,amssymb}
\usepackage{microtype}
\usepackage[prologue,table]{xcolor}
\usepackage{units}
\usepackage[binary-units]{siunitx}
\usepackage{subcaption}
\usepackage{overpic}
\usepackage{cancel}
\usepackage{wrapfig}
\usepackage{placeins}
\usepackage{empheq}

\usepackage{multirow}

\usepackage{url} 
\usepackage[utf8]{inputenc}

\def\commentType{0}

\ifnum\commentType=0
    \newcommandx{\customComment}[3]{}
    \newcommandx{\customTODO}[3]{}
\fi
\ifnum\commentType=1
    \usepackage[
    type=upperleft,
    unit=in
    ]{fgruler}
    \newcommandx{\customComment}[3]{\textcolor{#2}{\textsl{#1: #3}}}
    \newcommandx{\customTODO}[3]{\textcolor{#2}{\textsl{#1: #3}}}
\fi
\ifnum\commentType=2
    \usepackage[
    type=upperleft,
    unit=in
    ]{fgruler}
    \usepackage{pdfcomment}
    \newcommandx{\customComment}[3]{\pdfcomment[icon=Comment,opacity=0.5,color=#2,author=#1]{#3}}
    \newcommandx{\customTODO}[3]{\pdfcomment[icon=Note,opacity=0.5,color=#2,author=#1]{#3}}
\fi
\ifnum\commentType=3
    \usepackage[
    type=upperleft,
    unit=in
    ]{fgruler}
    \usepackage{pdfcomment} 
    \usepackage{todonotes} 
    \usepackage{silence}
    \newcommandx{\customComment}[3]{\todo[color=#2!40,size=\small]{\textbf{#1:} #3}}
    \newcommandx{\customTODO}[3]{\todo[color=#2!40,size=\small]{\textbf{#1:} #3}}
\fi

\let\originalleft\left 
\let\originalright\right 
\renewcommand{\left}{\mathopen{}\mathclose\bgroup\originalleft} 
\renewcommand{\right}{\aftergroup\egroup\originalright} 

\definecolor{amber}{rgb}{1.0, 0.49, 0.0}
\definecolor{darkgreen}{rgb}{0.0, 0.5, 0.0}
\newcommandx{\All}[1]{\customComment{All}{red}{#1}}
\newcommandx{\Marios}[1]{\customComment{Marios}{magenta}{#1}}
\newcommandx{\Ana}[1]{\customComment{Ana}{blue}{#1}}
\newcommandx{\Cengiz}[1]{\customComment{Cengiz}{amber}{#1}}
\newcommandx{\Thomas}[1]{\customComment{Thomas}{red}{#1}}

\newcommandx{\TODO}[1]{\customTODO{TODO}{red}{#1}}
\newcommandx{\MariosTODO}[1]{\customTODO{Marios TODO}{magenta}{#1}}
\newcommandx{\AnaTODO}[1]{\customTODO{Ana TODO}{blue}{#1}}
\newcommandx{\CengizTODO}[1]{\customTODO{Cengiz TODO}{amber}{#1}}
\newcommandx{\ThomasTODO}[1]{\customTODO{Thomas TODO}{darkgreen}{#1}}

\newcommand{\IGNORE}[1]{}

\newcommand{\REMOVE}[1]{} 
\newcommand{\ADD}[1]{#1} 

\def\figureautorefname{Figure}

\def\equationautorefname~#1\null{%
  Equation~(#1)\null
}

\newcommand{\autoreftable}[1]{\def\figureautorefname{Table}\autoref{#1}\def\figureautorefname{Figure}}

\def\approxprop{%
  \def\p{%
    \setbox0=\vbox{\hbox{$\propto$}}%
    \ht0=0.6ex \box0 }%
  \def\s{%
    \vbox{\hbox{$\sim$}}%
  }%
  \mathrel{\raisebox{0.7ex}{%
      \mbox{$\underset{\s}{\p}$}%
    }}%
}

\DeclareMathOperator{\sinc}{sinc}

\newcommand{\Sphere}{{\mathbb{S}^2}}

\newcommand{\nGaussians}{16}
\newcommand{\nGaussiansVar}{K}
\newcommand{\Normal}{\mathcal{N}}
\newcommand{\nSppPerIteration}{4}

\newcommand{\Diff}[1]{\,\mathrm{d}#1}

\newcommand{\PdfMC}{p}

\newcommand{\PdfBSDF}{p_{\bsdf}}

\newcommand{\toTangent}{\log}
\newcommand{\toWorld}{\exp}

\newcommand{\bsdfParams}{\phi}

\newcommand{\R}{\mathbb{R}}

\newcommand{\ssEmaDecay}{\eta}

\newcommand{\mixtureWeight}{\pi}
\newcommand{\weightPrior}{q}
\newcommand{\covPrior}{B}
\newcommand{\denomPrior}{a}

\newcommand{\diag}{\mathop{\mathrm{diag}}}

\newcommand{\dirt}{\nu}
\newcommand{\dir}{\omega}
\newcommand{\pos}{\mathbf{x}}
\newcommand{\diro}{\omega_\mathrm{o}}
\newcommand{\diri}{\omega_\mathrm{i}}

\newcommand{\normal}{{\mathrm{n}}}

\newcommand{\radiance}{L}
\newcommand{\inRadiance}{\radiance_{\mathrm{i}}}

\newcommand{\reflectedRadiance}{\radiance_{\mathrm{s}}}

\newcommand{\bsdf}{f_{\mathrm{s}}}

\newcommand{\PdfRadiance}{\PdfMC_{\inRadiance}}

\newcommand{\fsangle}{\gamma}

\newcommand{\PriorScale}{\beta}

\newcommand{\bigO}{\mathcal{O}}

\newcommand{\Bathroom}{\textsc{Bathroom}}
\newcommand{\Bedroom}{\textsc{Bedroom}}
\newcommand{\Bookshelf}{\textsc{Bookshelf}}
\newcommand{\Bottle}{\textsc{Bottle}}
\newcommand{\CornellBox}{\textsc{Cornell Box}}
\newcommand{\GlossyKitchen}{\textsc{Glossy Kitchen}}

\newcommand{\SwimmingPool}{\textsc{Swimming Pool}}

\newcommand{\Torus}{\textsc{Torus}}
\newcommand{\VeachDoor}{\textsc{Veach Door}}

\newcommand{\GlossyCornellBox}{\textsc{Yet Another Box}}

\newcommand{\WaterCaustic}{\textsc{Water Caustic}}
\newcommand{\Necklace}{\textsc{Necklace}}

\renewcommand{\fboxsep}{0pt}

\gdef\useCroppedImages{1}

\usepackage[export]{adjustbox}
\usepackage{calc}
\usepackage{tabularx}
\usepackage{tikz}
\usepackage{xstring}

\renewcommand{\fboxsep}{0pt}

\newlength{\beautyHeight}
\newlength{\beautyPixWidth}
\newlength{\beautyPixHeight}
\newlength{\insetvsep}
\gdef\useInsetA{0}
\gdef\useInsetB{0}
\gdef\useInsetC{0}

\newcommand{\setInset}[6]{%
    \expandafter\gdef\csname useInset#1\endcsname{1}%
    \expandafter\gdef\csname inset#1Color\endcsname{#2}%
    \expandafter\gdef\csname crop#1X\endcsname{#3}%
    \expandafter\gdef\csname crop#1Y\endcsname{#4}%
    \expandafter\gdef\csname crop#1W\endcsname{#5}%
    \expandafter\gdef\csname crop#1H\endcsname{#6}%
}

\newcommand{\unsetInset}[1]{%
    \expandafter\gdef\csname useInset#1\endcsname{0}%
}

\newcommand{\addBeautyCrop}[8]{%
    \pdfpxdimen=\dimexpr 1 in/72\relax
    \def\beauty{%
        \let\cropR\relax%
        \let\cropB\relax%
        \newlength\cropR%
        \newlength\cropB%
        \setlength\cropR{{#3 px}-{#5 px}-{#7 px}}%
        \setlength\cropB{{#4 px}-{#6 px}-{#8 px}}%
        \sbox0{\includegraphics[width=#2\textwidth,trim={#5px {\cropB} {\cropR} #6px},clip]{#1}}%
        \begin{tikzpicture}
            \node[anchor=north west,inner sep=0] at (0,0) {\usebox0};
            \begin{scope}[x=\wd0/#7, y=\ht0/#8]
            \if\useInsetA1{
                \draw[\insetAColor,thick] (\cropAX-#5,-\cropAY+#6) rectangle + (\cropAW,-\cropAH);
            }\fi
            \if\useInsetB1{
                \draw[\insetBColor,thick] (\cropBX-#5,-\cropBY+#6) rectangle + (\cropBW,-\cropBH);
            }\fi
            \if\useInsetC1{
                \draw[\insetCColor,thick] (\cropCX-#5,-\cropCY+#6) rectangle + (\cropCW,-\cropCH);
            }\fi
            \end{scope}
        \end{tikzpicture}
    }%
    \setlength\beautyHeight{\heightof{\beauty}}%
    \setlength\beautyPixWidth{#3px}%
    \setlength\beautyPixHeight{#4px}%
    \global\beautyHeight=\beautyHeight%
    \global\beautyPixWidth=\beautyPixWidth%
    \global\beautyPixHeight=\beautyPixHeight%
    \begin{adjustbox}{valign=t}
        \beauty{}
    \end{adjustbox}
}

\newcommand{\trimInset}[6]{%
    \let\cropR\relax%
    \let\cropB\relax%
    \newlength\cropR%
    \newlength\cropB%
    \setlength\cropR{{\beautyPixWidth}-{#3 px}-{#5 px}}%
    \setlength\cropB{{\beautyPixHeight}-{#4 px}-{#6 px}}%
    \color{#2}%
    \fbox{\includegraphics[width=1\linewidth,trim={{#3 px} {\cropB} {\cropR} {#4 px}},clip]{#1}}%
}

\newcommand{\addInset}[2]{%
    \color{#2}%
    \fbox{\includegraphics[width=1\linewidth]{#1}}%
}

\newcommand{\auxtimes}{x}
\newcommand{\auxplus}{+}
\newcommand{\auxspace}{ }

\newcommand{\addInsets}[1]{%
    \begin{adjustbox}{valign=t}
        \StrSubstitute{#1}{.}{-}[\baseFileName]
        \begin{adjustbox}{totalheight=1\beautyHeight,tabular={c}}
            \if\useInsetA1%
                \def\cropfile{\baseFileName-\cropAW\auxtimes\cropAH\auxplus\cropAX\auxplus\cropAY-crop}
                \if\cropInsets1
                    \immediate\write18{convert #1 -crop \cropAW\auxtimes\cropAH\auxplus\cropAX\auxplus\cropAY\auxspace -filter point -resize 800\% \cropfile.jpg}
                \fi
                \if\useCroppedImages1
                    \addInset{\cropfile.jpg}{\insetAColor}
                \else
                    \trimInset{#1}{\insetAColor}{\cropAX}{\cropAY}{\cropAW}{\cropAH}%
                \fi%
            \fi%
            \if\useInsetB1%
                \if\useInsetA1\\[\insetvsep]\fi%
                \def\cropfile{\baseFileName-\cropBW\auxtimes\cropBH\auxplus\cropBX\auxplus\cropBY-crop}
                \if\cropInsets1
                    \immediate\write18{convert #1 -crop \cropBW\auxtimes\cropBH\auxplus\cropBX\auxplus\cropBY\auxspace -filter point -resize 800\% \cropfile.jpg}
                \fi
                \if\useCroppedImages1
                    \addInset{\cropfile.jpg}{\insetBColor}
                \else
                    \trimInset{#1}{\insetBColor}{\cropBX}{\cropBY}{\cropBW}{\cropBH}%
                \fi%
            \fi%
            \if\useInsetC1%
                \if\useInsetB1\\[\insetvsep]\fi%
                \def\cropfile{\baseFileName-\cropCW\auxtimes\cropCH\auxplus\cropCX\auxplus\cropCY-crop}
                \if\cropInsets1
                    \immediate\write18{convert #1 -crop \cropCW\auxtimes\cropCH\auxplus\cropCX\auxplus\cropCY\auxspace -filter point -resize 800\% \cropfile.jpg}
                \fi
                \if\useCroppedImages1
                    \addInset{\cropfile.jpg}{\insetCColor}
                \else
                    \trimInset{#1}{\insetCColor}{\cropCX}{\cropCY}{\cropCW}{\cropCH}%
                \fi%
            \fi%
        \end{adjustbox}
    \end{adjustbox}
}

\definecolor{mathematicaBlue}{rgb}{0.38, 0.51, 0.71}
\definecolor{mathematicaOrange}{rgb}{0.88, 0.61, 0.14}
\definecolor{mathematicaGreen}{rgb}{0.56, 0.69, 0.19}
\definecolor{mathematicaRed}{rgb}{0.92,0.39, 0.21}
\definecolor{mathematicaPurple}{rgb}{0.53, 0.47, 0.7}

\definecolor{cvintegrand}{rgb}{1.0, 0.65, 0.0}
\definecolor{cvg}{rgb}{0.5, 0.0, 0.5}
\definecolor{cvG}{rgb}{0.67, 0.14, 0.19}

\definecolor{cvdifference}{rgb}{1.0, 0.65, 0.0}
\definecolor{cvpdf}{rgb}{0.5, 0.0, 0.5}

\begin{document}

\teaser{\centering
 \vspace{-8mm}
\begin{tabular}{cc}
  \begin{adjustbox}{valign=b}\begin{overpic}[height=5.15cm]{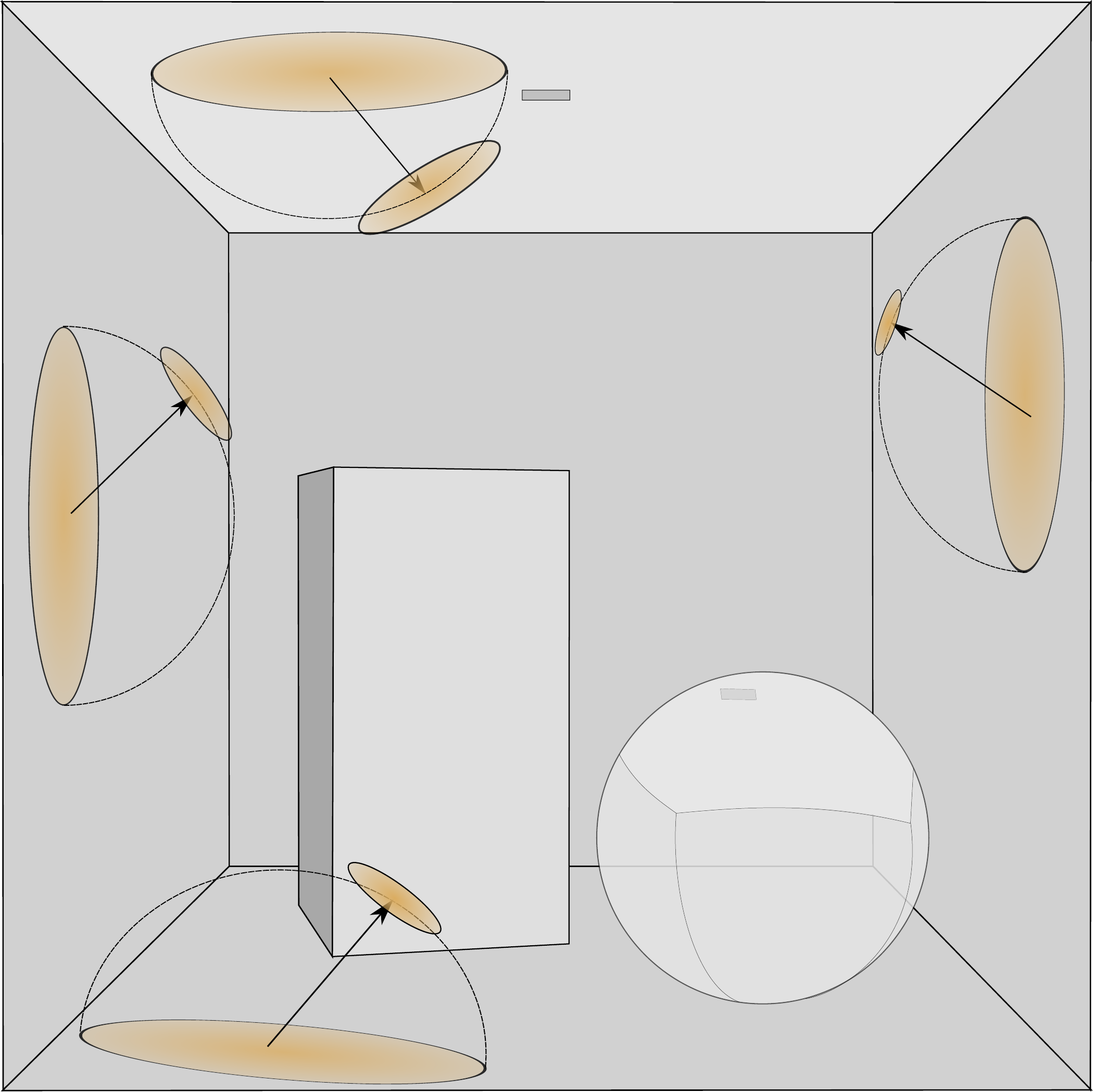}
    \put(20,100.5)  {\small\textbf{(a)} Illustration of SDMMs}
  \end{overpic}\end{adjustbox} &
\begin{adjustbox}{valign=b}\setlength{\fboxrule}{15pt}%
\setlength{\insetvsep}{30pt}%
\setlength{\tabcolsep}{-1.5pt}%
\renewcommand{\arraystretch}{1}%
\small%
\begin{tabular}{ccccc}
  & M\"uller et al. & \multicolumn{2}{c}{SDMM (Ours)} & \\
  \cmidrule(lr){2-2}
  \cmidrule(lr){3-4}
  \textbf{(b)} Rendering of \GlossyCornellBox{} & PPG & Radiance & Product & Reference \\
    \setInset{A}{red}{233}{133}{50}{60}%
    \setInset{B}{orange}{400}{283}{50}{60}%
    \addBeautyCrop{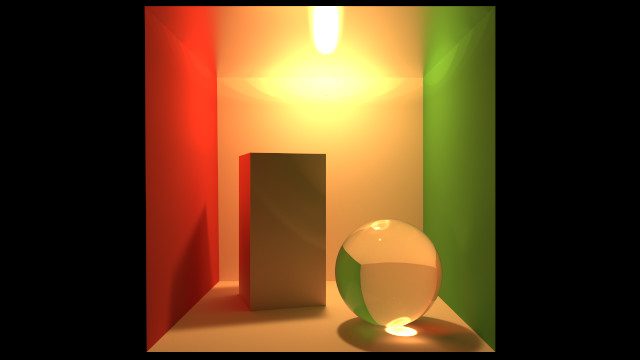}{0.2325}{640}{360}{147}{7}{346}{344} &
    \addInsets{images/fig-teaser/glossy-cbox-PPG_u.jpg} &
    \addInsets{images/fig-teaser/glossy-cbox-SDMM-radiance_u.jpg} &
    \addInsets{images/fig-teaser/glossy-cbox-SDMM-product_u.jpg} &
    \addInsets{images/fig-references/glossy-cbox-reference.jpg} \\
  \multicolumn{1}{r}{MAPE:} & 0.560 & 0.068 & \textbf{0.050} & \\
  \multicolumn{1}{r}{render time:} & 1.3m & 2.4m & 5.2m & \\
  \multicolumn{1}{r}{speedup vs.\ PPG at equal MAPE:} & 1.00$\times$ & \textbf{15.46$\times$} & 9.23$\times$ & \\
\end{tabular}
\end{adjustbox}
\end{tabular}
\caption{\label{fig:teaser}
  \textbf{(a)} Illustration of our $5$D spatio-directional mixture model (SDMM) that approximates incident radiance.
  \textbf{(b)} Rendering of a scene with difficult spatio-directionally varying illumination: a tiny light source is shining \emph{upwards} onto a glossy ceiling that indirectly illuminates the scene.
  We compare path guiding results with and without product sampling using an additional mixture that approximates the BSDF.\@
}}

\maketitle
\begin{abstract}
      We propose a learning-based method for
      light-path construction in path tracing algorithms, which iteratively optimizes and samples from
      what we refer to as spatio-directional Gaussian mixture models (SDMMs).
      In particular, we approximate incident radiance as an online-trained $5$D mixture that is accelerated by a $k$D-tree.
      Using the same framework, we approximate BSDFs as pre-trained $n$D mixtures, where $n$ is the number of BSDF parameters.
      Such an approach addresses two major challenges in path-guiding models.
      First, the $5$D radiance representation naturally captures correlation between the spatial and directional dimensions.
      Such correlations are present in e.g.\ parallax and caustics.
      Second, by using a tangent-space parameterization of Gaussians, our spatio-directional mixtures can perform approximate product sampling with arbitrarily oriented BSDFs.
      Existing models are only able to do this by either foregoing anisotropy of the mixture components or by representing the radiance field in local (normal aligned) coordinates, which both make the radiance field more difficult to learn.
      An additional benefit of the tangent-space parameterization is that each individual Gaussian is mapped to the solid sphere with low distortion near its center of mass.
      Our method performs especially well on scenes with small, localized luminaires that induce high spatio-directional correlation in the incident radiance.


\end{abstract}

\section{Introduction}%
\label{sec:introduction}%
\footnotetext{\small{\footnotemark Research completed prior to joining Facebook.}}

\setcounter{footnote}{0}
Path tracing is used to synthesize photorealistic images in a wide variety of settings, such as movie production, product and architecture visualization, and, more recently, video games.
The efficiency of path tracing is a product of two factors: (i) the rate at which paths are traced and (ii) how well the distribution of traced paths matches the actual distribution of light throughout the virtual scene.
The \emph{key challenge} is thus to devise an algorithm that accurately samples the distribution of light while at the same time remaining performant.
Adaptive importance sampling techniques (usually referred to as ``path guiding'' in the specific context of rendering) present a promising approach to this problem -- instead of relying on heuristics, they use information gathered during rendering to learn a sampling distribution which better suits the scene at hand.

Our method belongs to the class of path-guiding techniques which represent factors of the rendering integrand using mixture models.
In contrast to prior works which use $2$D directional mixture models~\cite{Vorba:2014:OnlineLearningPMMinLTS,herholz2016}, our method includes additional parameters of the rendering equation as further dimensions of the mixture model.
Specifically, we propose representing
\begin{itemize}
    \item incident radiance ($\inRadiance$) using a $5$D Gaussian mixture model, and
    \item BSDFs ($\bsdf$) using $n$D Gaussian mixture models,
\end{itemize}
where $n$ is the number of parameters of the BSDF.\@

The first benefit of such an aproach is that a $5$D mixture model naturally captures spatio-directional correlation (e.g.\ parallax) of the incident radiance $\inRadiance(\diri, \pos)$.
While recent work focuses on parallax as a specific instance of spatio-directional correlation, it effetively demonstrates the importance of accurately capturing it during path tracing~\cite{ruppert2020robust}.
Additionally, we found $n$D mixture models to be well-suited for compactly representing parametric BSDFs, as they both vary smoothly as a function of their parameters in most cases~\cite{Herholz:2018}.
Such a parametric BSDF representation is desirable, because it avoids the need of pre-computing multiple distinct BSDF models for every encountered material configuration.
Lastly, after both models have been trained, we can perform closed-form \emph{product sampling}.

In this paper, we describe solutions to overcome three challenges that would otherwise make the use of higher-dimensional mixture models difficult:

\paragraph*{Product sampling with mixed-orientation Gaussians.}
In order to compute the product of Gaussian distributions, they must live in the same coordinates.
Herholz et al.~\cite{herholz2016} achieve this by parameterizing incident radiance $\inRadiance$ as well as the BSDFs $\bsdf$ as $2$D directional distributions in the shading frame, where the surface normal always points towards ${(0,0,1)}$.
This approach is not possible for us, because $\inRadiance$ is approximated by a global $5$D spatio-directional mixture model, where all mixture components must share the same (global) coordinate frame.
For product sampling, we must therefore rotate the mixture into the shading frame on-the-fly for every sampling decision.
Such a rotation inherently introduces distortion due to the non-linear change of variables induced by changing the $2$D parameterization of the solid sphere.
To minimize this distortion near the center of mass of each Gaussian, we (i) adopt a tangent-space parameterization~\cite{pennec2006intrinsic,BMVC.28.91} as a substitute for commonly used parameterizations (e.g.\ cylindrical), and (ii) we propose to transform covariance matrices via a first-order Taylor approximation of the change of variables to further reduce the remaining distortion.

\paragraph*{Training of conditional distributions.}
In path guiding, we are interested in $2$D importance-sampling distributions.
Thus, when representing incident radiance $\inRadiance(\diri, \pos)$ as a $5$D distribution $\PdfMC(\diri, \pos)$, we ultimately only make use of the $2$D \emph{conditioned} distribution $\PdfMC(\diri \,|\, \pos)$.
The same holds true for BSDFs.
Unfortunately, the optimization of conditional distributions is highly non-linear and difficult~\cite{10.1145/1102351.1102446}.
Instead of attacking this difficulty head-on, we side-step it by optimizing our high-dimensional \emph{joint} distributions using the expectation maximization (EM) algorithm and only conditioning on-the-fly for each sampling decision.

\paragraph*{Scaling towards thousands of mixture components.}
The aforementioned conditioning is the conceptual equivalent to looking up a $2$D distribution in a higher-dimensional cache such as done by Vorba et al.~\cite{Vorba:2014:OnlineLearningPMMinLTS}.
When done with brute force, its cost scales linearly with the number of mixture components.
It is therefore an expensive operation that needs acceleration through an appropriate data structure, especially when the high-dimensional mixture consists of thousands of components, which is needed to represent a function as complicated as the incident radiance.
We use a continually evolving $k$D-tree over the conditional dimensions (e.g.\ the 3 spatial dimensions of incident radiance) to accelerate the on-the-fly conditioning operation to practical speed.

To summarize, the contributions of this paper include
\begin{itemize}
    \item the use of spatio-directional mixture models for approximating the incident radiance and BSDFs in product path guiding, which is made practical by
    \item a directional tangent-space parameterization that admits low distortion under rotation, and
    \item a $k$D-tree acceleration data structure for efficient on-the-fly conditioning and EM training.
\end{itemize}

\begin{figure}
    \centering
    \begin{subfigure}{.5\linewidth}
        \centering
        \includegraphics[width=\linewidth]{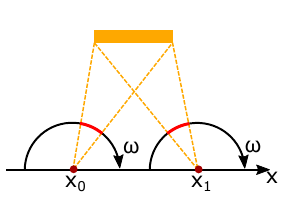}
        \caption{Scene setup}
        \label{fig:cboxggt}
    \end{subfigure}%
    \begin{subfigure}{.5\linewidth}
        \centering
        \includegraphics[width=\linewidth]{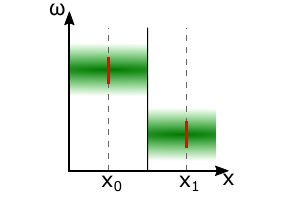}
        \caption{Vorba et al.~\cite{Vorba:2014:OnlineLearningPMMinLTS}}
        \label{fig:cboxgpath}
    \end{subfigure}
    \begin{subfigure}{.5\linewidth}
        \centering
        \includegraphics[width=\linewidth]{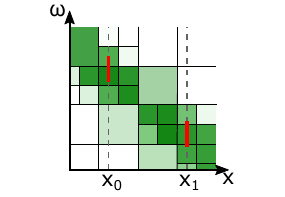}
        \caption{Practical Path Guiding~\cite{mueller2017practical}}
        \label{fig:cboxgppg}
    \end{subfigure}%
    \begin{subfigure}{.5\linewidth}
        \centering
        \includegraphics[width=\linewidth]{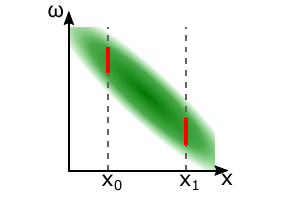}
        \caption{Ours}
        \label{fig:cboxgpmc}
    \end{subfigure}
    \caption{
        Comparison of our representation with previous work.
        In (a) we visualize a 2D incident radiance field with one spatial and one angular dimension $(x,\omega)$.
        The two points $x_0$ and $x_1$ are directly illuminated by an area light.
        (b) to (d) visualize how the incident radiance field is represented as a function of $x$ and $\omega$ in previous and our work.
        In this setup, (b) approximate the angular variation of incident radiance with two 1D GMMs at discrete spatial positions, (c) discretize both dimensions in a $2$D adaptive hierarchical data structure and finally, our method in (d) represents the entire spatio-angular domain with a $2$D GMM which captures the correlation between the dimensions.\@
    }\label{fig:2d-indicend-radiance}
\end{figure}

\section{Related Work}%
\label{sec:related-work}

Existing approaches to importance sampling the rendering integral range from analytic techniques, such as bi-directional path construction~\cite{Lafortune93}, Markov chain sampling~\cite{Veach:1997:Thesis}, multiple importance sampling~\cite{Veach:1995:MIS}, and BSDF models~\cite{Heitz:2014,Walter:2007}, to data-driven techniques, such as path guiding.

\paragraph*{Path guiding.}
``Path guiding'' techniques apply principles of adaptive importance sampling by iteratively
\begin{enumerate}
    \item tracing an initial set of paths,
    \item fitting an approximate radiance or product model to these paths,
    \item and then using the model to importance sample further paths.
\end{enumerate}
Such techniques were pioneered by Jensen~\cite{Jensen1995}, who rasterized a photon map into hemispherical grids for importance sampling, and Lafortune and Willems~\cite{Lafortune:1995:A5T}, who quantized radiance estimates into a spatio-directional $5$D tree that can be used both as a control variate and for importance sampling.
More recently, path-guiding techniques surged in popularity after Vorba et al.~\cite{Vorba:2014:OnlineLearningPMMinLTS} demonstrated that unidirectional path guiding can match bi-directional techniques in terms of sample quality---a boon to production rendering, which heavily relies on unidirectional tracing.
Consequently, many production renderers support path guiding nowadays~\cite{vorba19guiding}.

In our work, we adopt several breakthroughs of recent works on path guiding.
We perform reinforcement-learning-style iterated rendering~\cite{Dahm16,mueller2017practical}, where the rendering iterations are combined with an inverse-variance weighting~\cite{mueller19guiding}.
Furthermore, we use a spatial acceleration tree structure that adapts to the sample density~\cite{Herholz:2019}, and we train Gaussian mixture models to approximate incident radiance~\cite{Vorba:2014:OnlineLearningPMMinLTS} within the leaves of that spatial tree.
Lastly, we train another Gaussian mixture model to approximate BSDFs and we perform closed-form sampling of the product between the BSDF and the radiance models~\cite{herholz2016}.

\paragraph*{Spatio-directional effects.} Initial path-guiding approaches modeled spatio-directional correlations implicitly by subdividing space~\cite{Lafortune:1995:A5T,Jensen1995,Vorba:2014:OnlineLearningPMMinLTS,Hey:2002:ISH:584458.584476,Dahm16,mueller2017practical}.
More explicit modeling of such correlations came with primary sample space~\cite{Zheng:2019,mueller2019nis,Guo:2018} and path space approaches~\cite{reibold2018selective}, reprojection-based techniques~\cite{ruppert2020robust} as well as attempts at directly representing the full $5$D radiance or $7$D product using neural networks~\cite{mueller2019nis,mueller20neural}.
Our approach is most similar to the latter category---we also explicitly model the $5$D radiance field---but rather than using neural networks we use a mixture of $5$D Gaussians.
It is worth contrasting our data-driven $5$D mixtures to the reprojected $2$D mixtures of Ruppert et al.~\cite{ruppert2020robust}.
Reprojection constitutes an accurate closed-form solution when its underlying assumptions are valid, i.e.\ when the perceived origin is near-diffuse and the intermediate transport is spatially linear.
This is the case with emitters that are for example directly visible from the shading location or hidden behind flat mirrors or thin sheets of dielectrics.
\ADD{However, when these assumptions are violated, e.g.\ in caustics caused by curved specular reflectors, the data-driven approach is the more general one (see \autoref{fig:zoomed}).
Admittedly, such cases are rare in practice and do not result in complete failure---additional mixture components can compensate for erroneous reprojection---yet we believe the two approaches can be benefitially combined in the future.}

\paragraph*{Mixture models.}
Mixture models have a long and successful history for various uses in computer graphics.
Thus, for brevity, we will focus only on work that utilizes mixture models within the context of path guiding.
Hey et al.~\cite{Hey:2002:ISH:584458.584476} were the first to propose a mixture model for path guiding:\ a mixture of cone-shaped distributions seeded by a photon map.
Vorba et al.~\cite{Vorba:2014:OnlineLearningPMMinLTS} proposed to use a Gaussian mixture due to its ability to approximate sparse radiance estimates using the expectation maximization (EM) optimization algorithm.
They used an appproach similar to the general adaptive importance sampling algorithm called \emph{mixture population Monte Carlo}, by Capp\'e et al.~\cite{cappe:hal-00180669}.
Later approaches exploited the closed-form product of Gaussian mixtures to perform product sampling with the BSDF, where the BSDF is approximated by an auxiliary Gaussian~\cite{herholz2016} or skewed Gaussian~\cite{Herholz:2018} mixture.
Alternatively, mutivariate Gaussians can also parameterize the distribution of entire paths~\cite{reibold2018selective}, elegantly capturing spatio-directional correlations.
However, the high-dimensional space of paths is difficult to cover densely, constraining the approach to those paths whose interactions are near-specular and thereby have a low \emph{effective} dimensionality.

Similarly to Gaussian mixtures, mixtures of von Mises-Fisher distributions can also represent incident radiance, the BSDF (or phase function), and their product~\cite{Herholz:2019,ruppert2020robust}.
Von Mises-Fisher distributions have the key advantage that their domain is the solid sphere $\Sphere{}$ rather than Euclidean space, enabling their use without a spherical parameterization that could introduce unwanted distortion.
For path guiding, an additional crucial benefit of living in $\Sphere{}$ is that they can be trivially rotated, permitting product sampling between a world-space radiance representation and a local-space BSDF (or phase function) representation.

Our method seeks the same property for product sampling: the ability to rotate the mixture model freely in $\Sphere{}$.
However, we also desire the ability to model anisotropy with few mixture components as well as the ability to condition on additional Euclidean dimensions, e.g.\ the 3 spatial dimensions in the $5$D spatio-directional radiance distribution.
These requirements disqualify the isotropic von Mises-Fisher distributions, as well as other possible spherical candidates, such as the Bingham distribution~\cite{bingham} (undesirable symmetry) or anisotropic spherical Gaussians~\cite{Xu13sigasia} (difficult conditioning).
Instead, we rely on a \emph{tangent-space} Gaussian mixture model~\cite{pennec2006intrinsic,BMVC.28.91}, which we will describe in detail in \autoref{sec:tangent-space}.

\section{Methodology}%
\label{sec:methodology}

We are interested in using Monte Carlo integration to solve the reflection integral
\begin{align}
    \reflectedRadiance(\diro, \pos) = \int_\Sphere \inRadiance(\diri, \pos) \, \bsdf(\diri, \diro, \bsdfParams(\pos)) \cos(\fsangle) \Diff{\diri} \,,
\end{align}
where $\diro$ is the reflected direction, $\diri$ is the direction of incident radiance, $\fsangle$ is its angle with the surface normal, $\pos$ is the shading location, and $\bsdfParams(\pos)$ are the parameters of the BSDF at that location.

In order to improve the efficiency of said Monte Carlo integration, we would like to importance sample according to a probability density function (PDF) that is approximately proportional to the integrand, i.e.\ our goal is to find
\begin{align}
    \PdfMC(\diri \,|\, \diro,\pos) \propto \inRadiance(\diri,\pos) \, \bsdf(\diri,\diro,\bsdfParams(\pos)) \cos(\fsangle) \,.
    \label{eq:goal}
\end{align}

We approach this goal by decomposing it into two sub-goals: (i) learning the incident radiance $\inRadiance$ as a $5$D Gaussian mixture model and (ii) learning the BSDFs $\bsdf$ as $n$D Gaussian mixture models, where $n$ is the total number of dimensions of $\diri$, $\diro$, and $\bsdfParams$:
\begin{align}
    \PdfRadiance(\diri, \pos) &\propto \inRadiance(\diri,\pos) \, \\
    \PdfBSDF(\diri, \diro, \bsdfParams) &\propto \bsdf(\diri,\diro,\bsdfParams(\pos)) \cos(\fsangle) \, .
\end{align}

We then approximately sample according to \autoref{eq:goal} by conditioning the approximations $\PdfRadiance$ and $\PdfBSDF$ on $(\diro, \pos, \bsdfParams(\pos))$ and then computing their closed-form product distribution, i.e.
\begin{align}
    \PdfMC(\diri \,|\, \diro,\pos) \propto \PdfRadiance(\diri \,|\, \pos) \, \PdfBSDF(\diri \,|\, \diro, \bsdfParams(\pos)) \,.
\end{align}

Next, we will discuss each of these steps in detail.

\subsection{Tangent-Space Gaussian Mixtures}%
\label{sec:tangent-space}

\begin{figure}
    \vspace{-1mm}
    \begin{subfigure}{.5\linewidth}
        \centering
        \begin{overpic}[trim=10 0 280 0, clip, width=\linewidth]{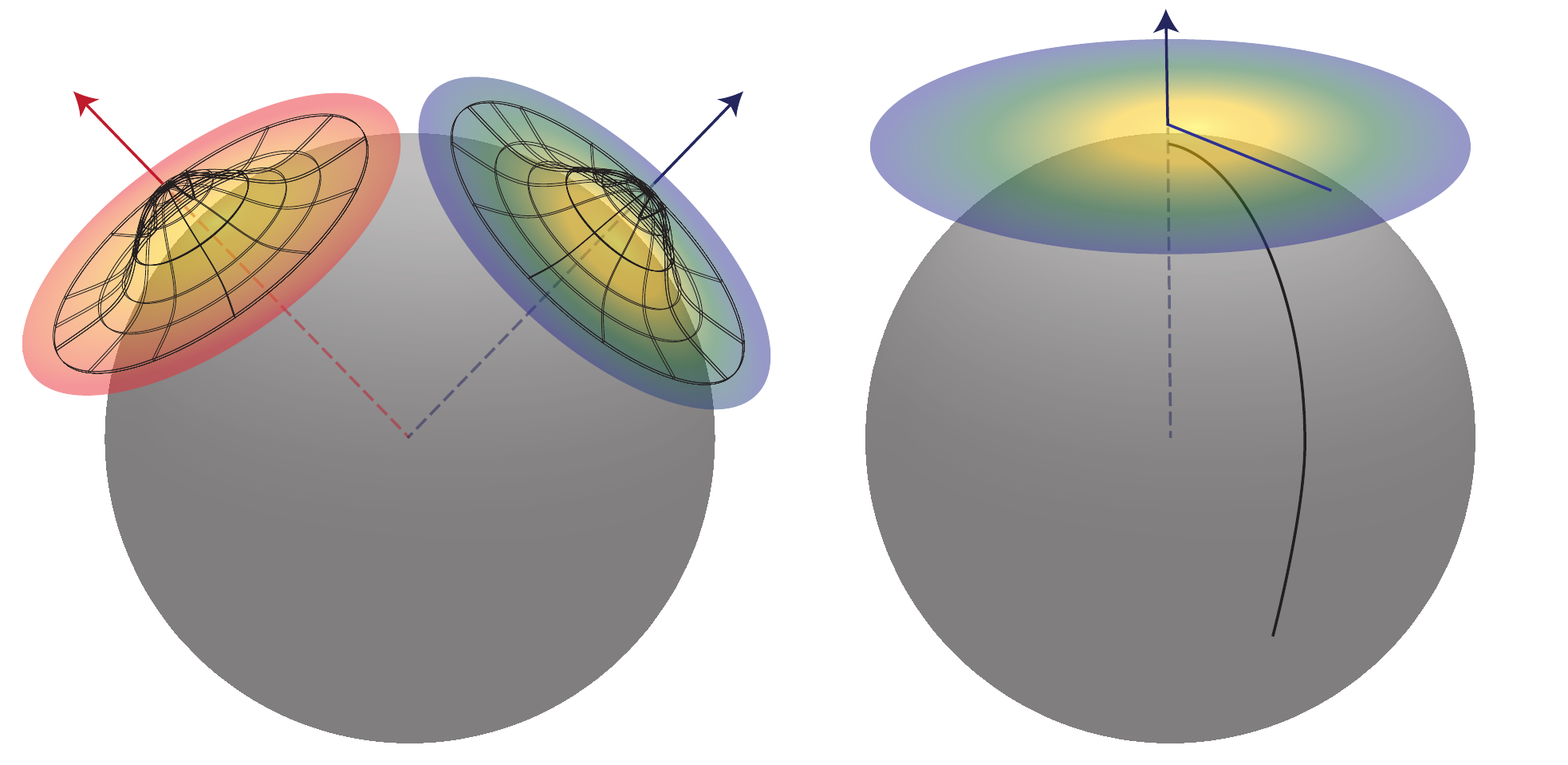}
            \put(15,90) {$\Normal(0, \Sigma_1)$}
            \put(60,90) {$\Normal(0, \Sigma_2)$}
            \put(30,50) {$\mu_1$}
            \put(63,50) {$\mu_2$}
        \end{overpic}
        \caption{Mixture setup}
        \label{fig:asdfgh}
    \end{subfigure}%
    \begin{subfigure}{.5\linewidth}
        \centering
        \begin{overpic}[trim=290 0 0 0, clip, width=\linewidth]{images/tangent-space.pdf}
            \put(49,85) {$0$}
            \put(73,75) {$\dirt$}
            \put(42,50) {$\mu$}
            \put(70,5) {$\dir$}
        \end{overpic}
        \caption{Tangent space parameterization}
        \label{fig:asdf}
    \end{subfigure}
    \caption{\label{fig:tangent-gaussians}
        \textbf{(a)} A 2-component example of a tangent-space Gaussian mixture model.
        Each Gaussian component is parameterized by a unit-length mean vector $\mu_k$, pointing to its position on the sphere, and a covariance matrix $\Sigma_k$ describing the Gaussian's shape in the corresponding \emph{tangent space} of the sphere.
        \textbf{(b)} The $\mu$-centered tangent space is a circular $2$D parameterization of the surface of the sphere. World-space directions $\dir$ are transformed to tangent-space directions $\dirt$ and back via $\dirt = \log_\mu(\dir)$ and $\dir = \exp_\mu(\dirt)$.
    }\vspace{-3mm}
\end{figure}

Since we represent incident radiance using a spatio-directional $5$D mixture model as opposed to caches that are associated with geometric surfaces, we are forced to adopt a global ``world-space'' parameterization of the incident radiance.
In order to compute product sampling between world-space radiance and a locally parameterized BSDF, we must be able to rotate the mixtures to the same coordinate frame.
To this end, we build upon spherical \emph{tangent-space} Gaussian mixture models~\cite{pennec2006intrinsic,BMVC.28.91}.
For example, in a tangent-space model on a sphere with $\nGaussiansVar$ mixture components, the $k$-th Gaussian is parameterized by a $3$D world-space unit-length mean vector ${\mu_k \in \Sphere}$ and a $2\!\times\!2$ tangent-space covariance matrix ${\Sigma_k \in \R^{2\times2}}$.
The mean vector $\mu_k$ \emph{determines} the tangent space that the covariance matrix $\Sigma_k$ lives in; see \autoref{fig:asdfgh} for an illustration.

Crucially, this representation can be rotated into any local shading frame for product sampling:\ one simply rotates the mean vectors $\mu_k$ to the local frame and \ADD{applies the azimuthal part of the rotation to the covariance matrices $\Sigma_k$}.

\paragraph*{Tangent spaces.}
We define the tangent space at some mean vector $\mu$, as a circular $2$D parameterization of the surface of the sphere~(see \autoref{fig:asdf}): one can map from the $\mu$-centered tangent space to the sphere and back, using the so-called $\toTangent$ and $\toWorld$ maps, which are each other's inverse:
\begin{align}
    \dirt = \toTangent_{\mu}(\dir) \,, \,\,\,\,\, \dir = \toWorld_{\mu}(\dirt) \,.
\end{align}

In the following we will assume that ${\dir \in \Sphere}$ is a direction vector and ${\dirt \in \R^2}$ is a coordinate in a tangent space with $\| \nu \| < \pi$.
In this case, we can use the \emph{azimuthal equidistant projection} to define our $\toTangent$ and $\toWorld$ maps, same as Simo-Serra et al.~\cite{BMVC.28.91}:
\begin{align}
    &\begin{gathered}
        \toTangent_{\mu}(\dir) = {\left(\frac{\dir^{\circlearrowleft}_x}{\sinc(\cos^{-1}(\dir^{\circlearrowleft}_z))} \,, \, \frac{\dir^{\circlearrowleft}_y}{\sinc(\cos^{-1}(\dir^{\circlearrowleft}_z))}\right)}^\intercal \,, \\
        \qquad \dir^{\circlearrowleft} = R_\mu \dir \, \,,
    \end{gathered} \label{eq:log} \\[10pt]
    &\toWorld_{\mu}(\dirt) = R_\mu^{-1} {\Big( \nu_u \sinc(\| \nu \|) , \, \nu_v \sinc(\| \nu \|) , \, \cos(\| \nu \|) \Big)}^\intercal \,,
    \label{eq:exp}
\end{align}
where $R_\mu$ is the rotation matrix that rotates $\mu$ to ${(0, 0, 1)}$ and $\sinc$ is the \emph{unnormalized} sinc function.

The crucial property that justifies the use of exponential and logarithm maps is that the length of the shortest path (i.e. shortest geodesic) connecting $\mu$ with any $\omega$ along the manifold (e.g. sphere) equals the distance between the origin of the tangent space at $\mu$ and the corresponding $\nu$ \footnote{Note that, for spheres, this property holds as long as $\| \nu \| < \pi$, or, more generally, $\nu$ does not cross the \emph{cut locus} at $\mu$}.
This property enables the optimization of tangent-space Gaussians using the expectation maximization (EM) algorithm~\cite{BMVC.28.91}, which we will describe in \autoref{sec:map-em-with-tangent-spaces}.

\paragraph*{Density evaluation.}
Given the parameters $\mu$ and $\Sigma$ of a spherical tangent-space Gaussian, its density \emph{in tangent space} is the regular bi-variate Gaussian density, centered at the tangent space's origin:
\begin{align}
    \PdfMC^t(\dirt) := e^{-\frac{1}{2} \dirt^\intercal \Sigma^{-1} \dirt} \big/ \left(2\pi\sqrt{\det{\Sigma}}\right) \,.
    \label{eq:bi-variate}
\end{align}
Its density with respect to the \emph{solid-angle measure}, which we need for importance sampling of directions in path tracing, is in turn obtained through a change of variables:
\begin{align}
    \PdfMC^\Omega(\dir) = \PdfMC^t(\dirt) \sqrt{|\det{G}|} \,,
\end{align}
where $G$ is the matrix representation of the metric tensor of the tangent space.
Specifically, it is the $2\!\times\!2$ matrix $G = J^\intercal J$, where $J$ is the $3\!\times\!2$ Jacobian matrix of the the exponential map, $\toWorld_{\mu}$, evaluated at $\nu$.\footnote{The use of the metric tensor in a change of variables generalizes the well-known use of the absolute determinant of the Jacobian matrix to the case of transforming between manifolds; in our case, we are mapping from a $2$D tangent space to $3$D unit vectors.}
We provide the closed form of the Jacobian matrix in \autoref{app:jacobians}.

Thus, the solid-angle density of a \emph{mixture} of $\nGaussiansVar$ tangent-space Gaussians is
\renewcommand{\fboxsep}{4pt}
\begin{empheq}[box=\fbox]{equation}
\begin{split}
    \PdfMC(\dir) := \sum_{k=1}^\nGaussiansVar \mixtureWeight_k\, \PdfMC_k^\Omega(\dir) \,,
    \label{eq:solidangle-pdf}
\end{split}
\end{empheq}
\renewcommand{\fboxsep}{0pt}%
where $\mixtureWeight_k$ is the weight of the $k$-th mixture component, such that ${\sum_k \mixtureWeight_k = 1}$, and $\PdfMC_k^\Omega(\dir)$ is its solid-angle density.

\paragraph*{Sampling.}
Sampling of a tangent-space mixture is simple compared to evaluating its density.
It consists of three steps: (i) sample the mixture component index $k$ proportional to $\mixtureWeight_k$, then (ii) sample the tangent-space direction $\dirt \propto \PdfMC_k^t(\dirt)$, and lastly (iii) convert the sampled $\dirt$ to world space by evaluating $\dir = \toWorld_{\mu_k}(\dirt)$.

Special care must be taken in step (ii), since the bijectivity of the azimuthal equidistant projection breaks down at the radius $\pi$: due to the equidistance property, the projection reaches the antipodal point of $\mu_k$ at the radius $\pi$.
Therefore, for the purpose of Monte Carlo integration using the density defined in \autoref{eq:solidangle-pdf}, if the sampled $\dirt$ lies outside the radius $\pi$, the sample must be discarded.

Note that all bounded parameterizations of the sphere (e.g.\ cylindrical coordinates) require discarding samples that lie outisde of their domain.
In fact, the tangent space formulation has an advantage in this regard, because the center of mass of each Gaussian is located in the center of its tangent space, leading to only a vanishingly small number of discarded samples in practice (on average $0.021$\%).

\paragraph*{Additional dimensions.}
When we want a tangent-space Gaussian mixture to admit additional, Euclidean dimensions, e.g.\ the position $\pos$ when approximating incident radiance as $\PdfRadiance(\diri, \pos) \approxprop \inRadiance(\diri,\pos)$, the above equations stay almost exactly the same.

In this case, \autoref{eq:bi-variate} becomes the general multi-variate Gaussian distribution, and the exponential and logarithm maps of the Euclidian dimensions are defined as \emph{translations} by their mean vector.
For example, the $3$D world-space position is mapped to its tangent space as $\toTangent_\mu{(\pos)} := \pos - \mu^\pos$, where $\mu^\pos$ is the mean of the Gaussian distribution in $3$D world space.
By representing Euclidean dimensions in their own tangent space, all dimensions---regardless of whether they are spherical or Euclidean---can be treated using the same tangent-space formulae.

\subsection{Learning of Conditional Mixture Models using EM}%
\label{sec:map-em-with-tangent-spaces}

We are ultimately interested in obtaining the conditional $2$D importance-sampling distributions $\PdfRadiance(\diri \,|\, \pos)$ and $\PdfBSDF(\diri \,|\, \diro, \bsdfParams)$.
However, directly optimizing the conditional distributions turns out to be difficult~\cite{10.1145/1102351.1102446}.
Therefore, we instead optimize the corresponding joint distributions $\PdfRadiance(\diri, \pos)$ and $\PdfBSDF(\diri, \diro, \bsdfParams)$, and subsequently condition them on-the-fly during rendering using standard closed-form Gaussian conditioning.
To optimize the joint distributions, we utilize the well established expectation maximization (EM) algorithm.
The following paragraphs briefly introduce EM and then focus on our adaptations to make it compatible with tangent-space Gaussian mixtures.

\paragraph*{Mini-batch expectation maximization.}
We use the mini-batch variant of EM~\cite{Dodik2020PathGU, Nguyen_2020} due to its online training capability\footnote{We discuss an alternative option, stepwise EM~\cite{Capp__2009, cappe:hal-00532968, Vorba:2014:OnlineLearningPMMinLTS}, in \autoref{sec:discussion}.}.
Mini-batch EM operates by repeatedly sampling a mini-batch of $M$ samples from a Gaussian mixture parameterized by $(\mixtureWeight_k, \mu_k, \Sigma_k)$ and then computing from that mini-batch a triplet of sufficient statistics $(S_k^{(0)}, S_k^{(1)}, S_k^{(2)})$ for each mixture component $k$.
From these sufficient statistics we then compute the updated mixture parameters, $(\hat{\mixtureWeight}_k, \hat{\mu}_k, \hat{\Sigma}_k)$.

Given a mini-batch of samples $x_1, \ldots, x_M \in \R^D$ with corresponding Monte Carlo weights $w_1, \ldots, w_M \in \R$, the sufficient statistics are computed as weighted sums
\begin{align}
    S_k^{(0)} = \frac{\sum_i^M w_i \mathcal{S}_{ki}}{\sum_i^M w_i} \,, \,\,\,\,\,
    S_k^{(1)} &= \frac{\sum_i^M w_i \mathcal{S}_{ki} x_i}{\sum_i^M w_i} \,, \,\,\,\,\,
    S_k^{(2)} = \frac{\sum_i^M w_i \mathcal{S}_{ki} x_i x_i^\intercal}{\sum_i^M w_i} \,, \nonumber \\
    \mathcal{S}_{ki} &:= \frac{\mixtureWeight_k \PdfMC_k^t(x_i)}{\PdfMC^t(x_i)} \,,
    \label{eq:em-ss-computation}
\end{align}
and, thereafter, the updated parameters $(\hat{\mixtureWeight}_k, \hat{\mu}_k, \hat{\Sigma}_k)$ are computed as
\begin{align}
    \hat{\mixtureWeight}_k = \frac{S_k^{(0)}}{\sum_i^\nGaussiansVar S_i^{(0)}} \,, \,\,\,\,\,
    \hat{\mu}_k = \frac{S_k^{(1)}}{S_k^{(0)}} \,, \,\,\,\,\,
    \hat{\Sigma}_k = \frac{S_k^{(2)}}{S_k^{(0)}} - \hat{\mu}_k \hat{\mu}_k^\intercal \,.
    \label{eq:em-params-from-ss}
\end{align}
Unfortunately, a derivation of the above equations is beyond the scope of this paper.
We refer to detailed descriptions of mini-batch EM based on sufficient statistics~\cite{Dodik2020PathGU, Nguyen_2020}.

\paragraph*{Sufficient statistics in tangent spaces.}
When the mixture components are defined in tangent space, special care must be taken when computing the first- and second-moment sufficient statistics $S_k^{(1)}$ and $S_k^{(2)}$.
Given a mini-batch of \emph{direction vectors} $\dir_i \in \Sphere, i \in \{1,\ldots, M\}$, the first-moment sufficient statistic $S_k^{(1)}$ of the $k$-th component must be computed in the corresponding $\mu_k$-centered tangent-space coordinates $\dirt_i =  \toTangent_{\mu_k}(\dir_i)$ and, subsequently, the updated mean vector $\hat{\mu}_k$ must be computed by translating back from the $\mu_k$-centered tangent-space to world space:
\begin{align}
    S_k^{(1)} = \frac{\sum_i^M w_i \mathcal{S}_{ki} \dirt_i}{\sum_i^M w_i} \,, \,\,\,\,\,
    \hat{\mu}_k = \toWorld_{\mu_k}\left( S_k^{(1)} \big/ S_k^{(0)} \right) \,.
    \label{eq:tangent-space-mu-update}
\end{align}
Only after the updated mean vector $\hat{\mu}_k$ has been computed can the second-moment sufficient statistic $S_k^{(2)}$ be computed.
This is the case, because the statistic $S_k^{(2)}$ determines the updated covariance matrix $\hat{\Sigma}_k$, which must be centered around $\hat{\mu}_k$, not $\mu_k$.
Formally:
\begin{align}
    S_k^{(2)} = \frac{\sum_i^M w_i \mathcal{S}_{ki}  \hat{\dirt}_i \hat{\dirt}_i^\intercal}{\sum_i^M w_i} \,, \,\,\,\,\,
    \hat{\Sigma}_k = S_k^{(2)} \big/ S_k^{(0)}
    \label{eq:tangent-space-sigma-update}
\end{align}
where $\hat{\dirt}_i = \toTangent_{\hat{\mu}_k}(\dir_i)$ and $\mathcal{S}_{ki}$ must use updated densities w.r.t.\ to the $\hat{\mu}_k$-centered tangent space.
Note, that the subtraction of $\hat{\mu}_k \hat{\mu}_k^\intercal$ is missing (c.f.~\autoref{eq:em-params-from-ss}), because $S_k^{(2)}$ was computed in the already $\hat{\mu}_k$-centered tangent space.

As previously stated in \autoref{sec:tangent-space}, additional Euclidean dimensions, such as the $3$D position $\pos$, can also be treated with the new formulae if the $\toTangent$ and $\toWorld$ maps map them to their respective tangent space by translating them by their mean vector.

\begin{figure}
    \begin{subfigure}{.5\linewidth}
        \centering
	\includegraphics[width=\linewidth]{images/glossy-cbox-flipped-light}
        \caption{Single mixture}
        \label{fig:asdfgh1}
    \end{subfigure}%
    \begin{subfigure}{.5\linewidth}
        \centering
        \includegraphics[width=\linewidth]{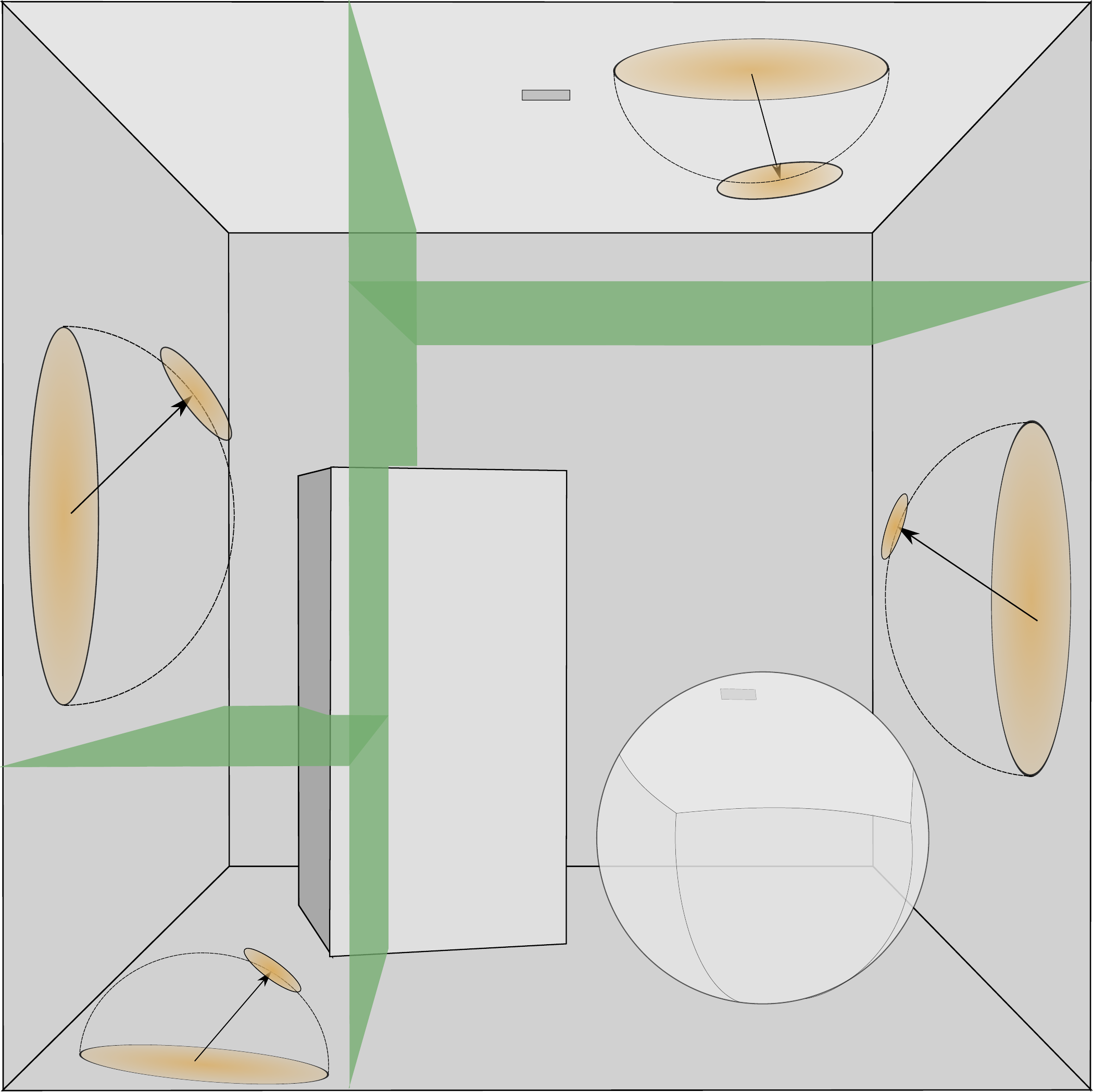}
        \caption{Mixture partitioned by $k$D-tree}
        \label{fig:kd-tree-right}
    \end{subfigure}
    \caption{
        Illustrations of the spatial components of our model. In (a) we visualize a small fraction of the mixture components within a single model and their spatial overlap.
        Due to the high amount of overlap, the computational requirements of updating and querying the model can be linear to the total number of components.
        By using a $k$D-tree spatial subdivision scheme (b) we can significantly reduce these computational requirements while still capturing correlations within a leaf node.
    }\label{fig:kd-tree}
    \vspace{-3mm}
\end{figure}

\subsection{Efficient Conditioning and EM}%
\label{sec:s-tree}

After the joint distribution $\PdfRadiance(\diri, \pos)$ has been optimized using EM, we seek to efficiently condition it on $\pos$ on-the-fly during rendering to be able to importance sample $\PdfRadiance(\diri \,|\, \pos)$.
However, the computational cost of conditioning is linear in the number of mixture components, and our $5$D Gaussian mixtures require \emph{thousands} of components to cover the incident radiance field well.
Thus, when done na\"ively for \emph{all} thousands of components, conditioning is exceedingly expensive.

The key observation that makes conditioning tractable is that only those mixture components whose spatial mean component $\mu_k^\pos$ is in close proximity to $\pos$ meaningfully affect the conditional distribution $\PdfRadiance(\diri \,|\, \pos)$.
Conditioning can thus be accelerated by only operating on a local neighborhood of mixture components.

The EM algorithm, like conditioning, also suffers from an $\bigO(\nGaussiansVar)$ cost and can therefore also be accelerated by operating independently on local neighborhoods of Gaussians.
At first, it may seem like restricting EM to local neighborhoods might destroy the desirable properties of spatio-directional mixtures, but this is a largely incorrect notion.
It merely prevents individual mixture components from covering large spatial regions, which is rarely appropriate in high-fidelity virtual scenes.
Most importantly, it does \emph{not} remove the key ability of spatio-directional mixtures to accurately model spatio-directional correlations such as parallax \emph{within} their neighborhoods.

\begin{figure}
    \vspace{-2mm}
    \setlength{\fboxrule}{10pt}%
\setlength{\insetvsep}{20pt}%
\setlength{\tabcolsep}{-1.5pt}%
\renewcommand{\arraystretch}{1}%
\small%
\begin{tabular}{cccc}
  & \multicolumn{2}{c}{SDMM} & DMM \\
  \cmidrule(lr){2-3}
  \cmidrule(lr){4-4}
   & Global & $k$D-tree & $k$D-tree \\
    \setInset{A}{red}{233}{133}{40}{60}%
    \setInset{B}{orange}{443}{225}{40}{60}%
    \addBeautyCrop{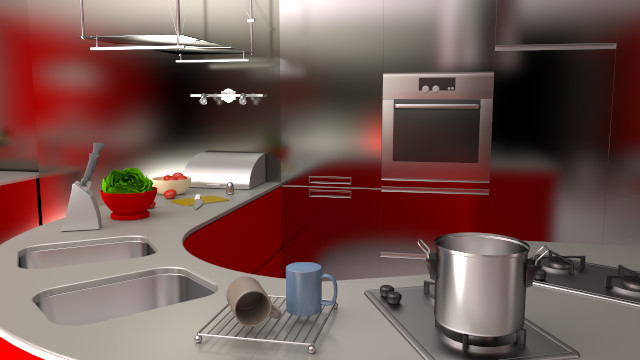}{0.218}{640}{360}{200}{0}{315}{360} &
    \addInsets{images/fig-mixture-comparison/glossy-kitchen-Single-mixture_u.jpg} &
    \addInsets{images/fig-mixture-comparison/glossy-kitchen-SDMM-radiance_u.jpg} &
    \addInsets{images/fig-mixture-comparison/glossy-kitchen-DMM-radiance_u.jpg} \\
  \multicolumn{1}{r}{MAPE:} & \textbf{0.219} & 0.225 & 0.427 \\
  \multicolumn{1}{r}{render time:} & 15h & 3.0m & 1.9m \\
\end{tabular}

    \vspace{-1mm}
    \caption{
        \ADD{Subdividing our spatio-directional mixture model by a $k$D-tree (middle) unlocks similarly low noise as using a single global mixture (left) at a fraction of the cost---one that is comparable to placing purely directional mixtures in the same $k$D-tree (right).
        All mixture models use a total of roughly 16000 components and were rendered with 1024spp.
        Additional comparisons of spatio-directional versus directional mixtures are available in the supplementary material.}
    }\label{fig:mixture-comparison}
\end{figure}

To partition our $5$D Gaussian mixture components into local neighborhoods, we adopt the spatial $k$D-tree partitioning scheme proposed by Herholz et al.~\cite{Herholz:2019}.
We subdivide and collapse the $k$D-tree according to the number of observed samples within each leaf---same as Herholz et al.---and \emph{within} each leaf, we place $\nGaussians$ of our $5$D mixture components, as illustrated in \autoref{fig:kd-tree-right}.
Given a query position $\pos$, conditioning is thus only performed on the $\nGaussians$ Gaussians found within the $k$D-tree's leaf node at $\pos$.
Likewise, given a mini-batch of spatio-directional samples for EM training, the samples are partitioned into their respective $k$D-tree leaves and mini-batch EM is applied to each leaf independently.

\ADD{In \autoref{fig:mixture-comparison} we demonstrate that using the $k$D-tree yields the desired performance benefits without compromising on the quality benefits of the spatio-directional radiance representation.}

\begin{figure*}
    \centering
    \begin{tabular}{r}
        \rotatebox{90}{\hspace{0.5cm}World space}\\[1cm]
        \rotatebox{90}{\hspace{0cm}Tangent space}
    \end{tabular}%
    \begin{subfigure}{.2\linewidth}
        \centering
        \includegraphics[width=\linewidth]{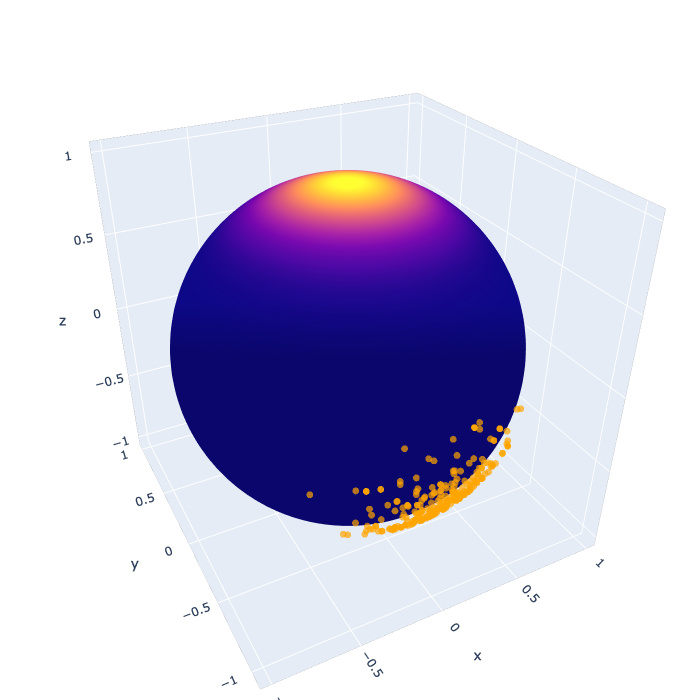}\\
        \includegraphics[width=\linewidth]{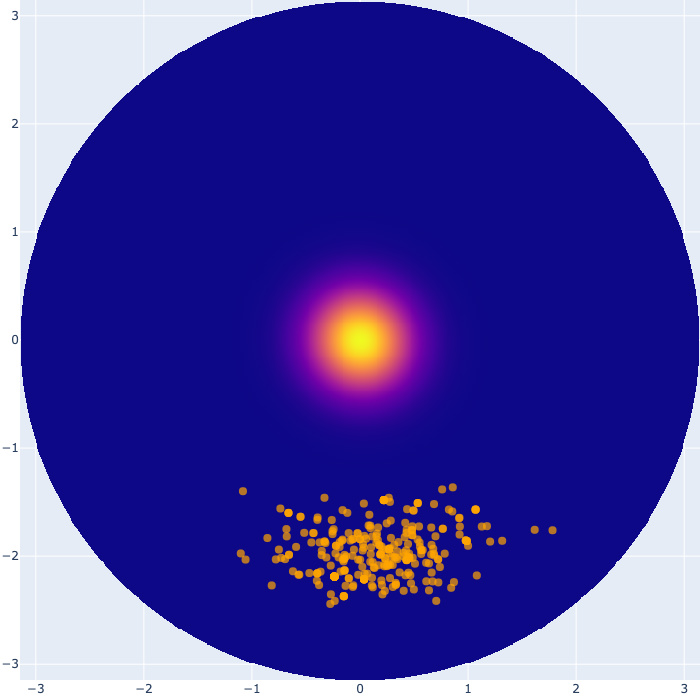}
        \caption{Before EM}
        \label{fig:cov-init}
    \end{subfigure}%
    \begin{subfigure}{.2\linewidth}
        \centering
        \includegraphics[width=\linewidth]{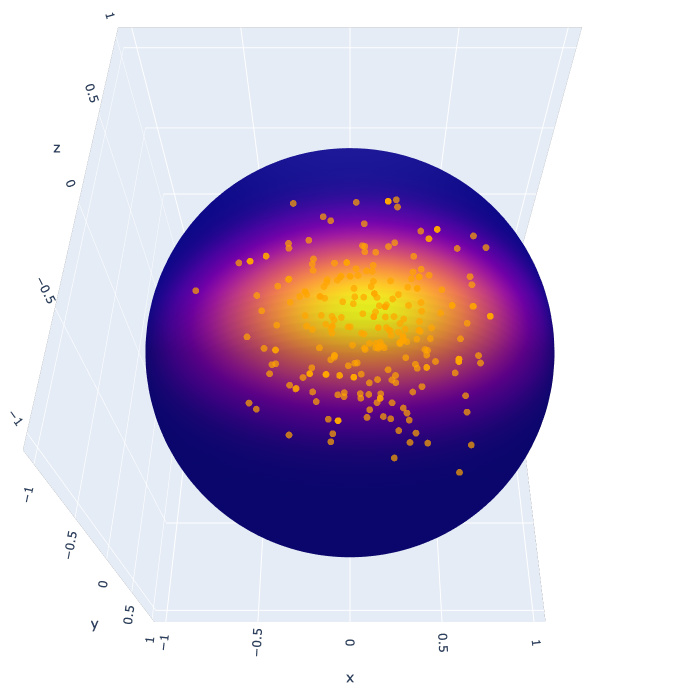}\\
        \includegraphics[width=\linewidth]{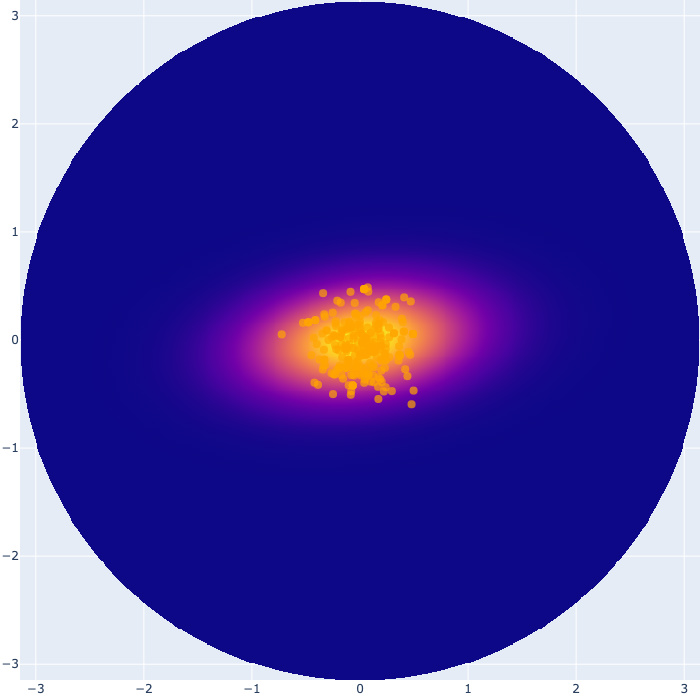}
        \caption{After EM (na\"{\i}ve $\Sigma$)}
        \label{fig:cov-naive}
    \end{subfigure}%
    \begin{subfigure}{.2\linewidth}
        \centering
        \includegraphics[width=\linewidth]{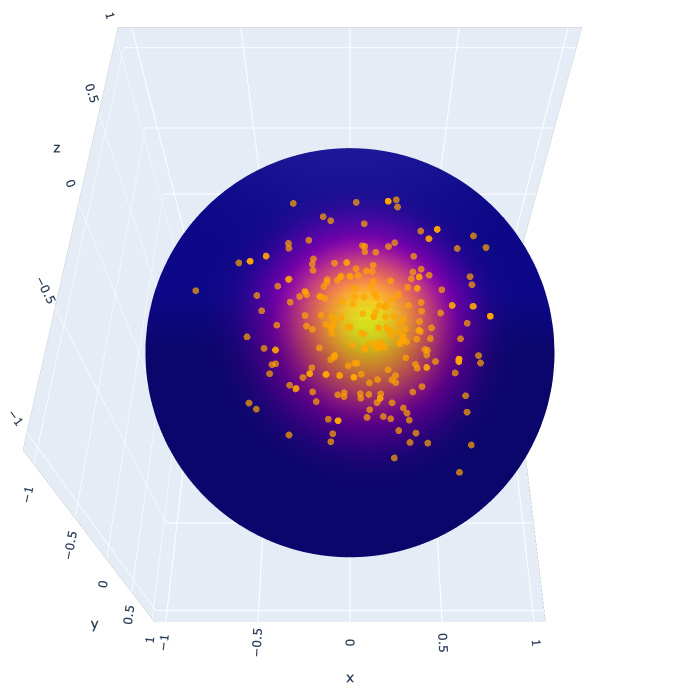}\\
        \includegraphics[width=\linewidth]{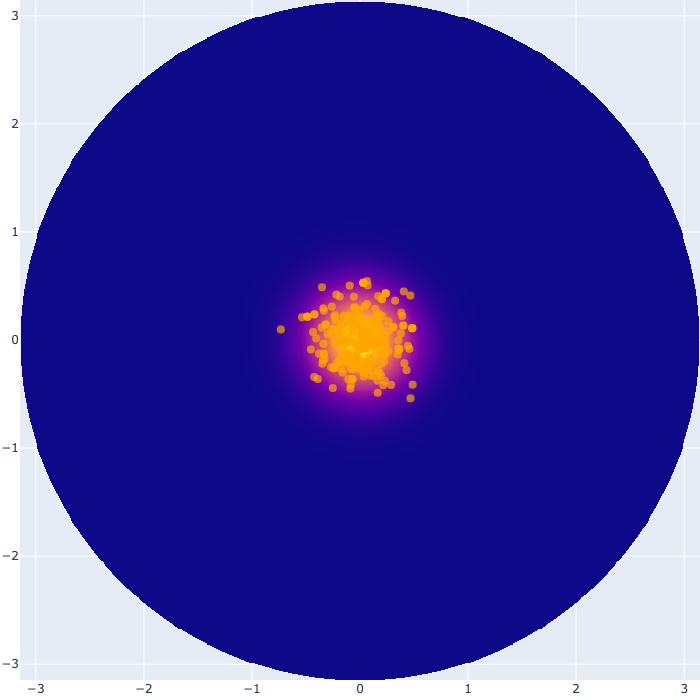}
        \caption{After EM (accurate $\Sigma$)}
        \label{fig:cov-accurate}
    \end{subfigure}%
    \begin{subfigure}{.2\linewidth}
        \centering
        \includegraphics[width=\linewidth]{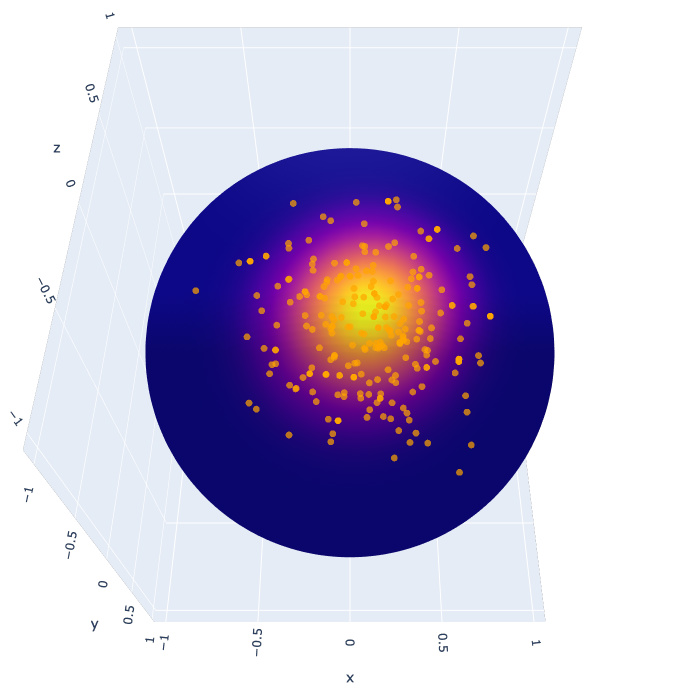}\\
        \includegraphics[width=\linewidth]{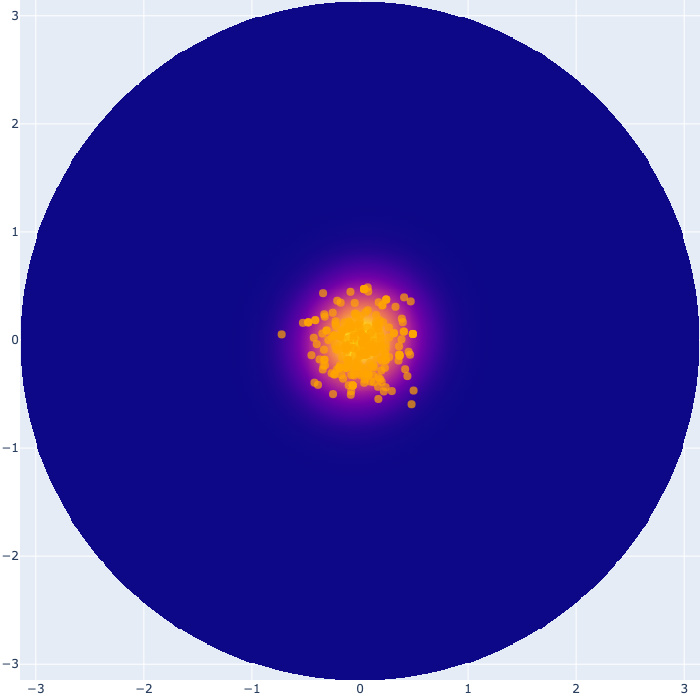}
        \caption{After EM (first-order $\Sigma$)}
        \label{fig:cov-first-order}
    \end{subfigure}

    \caption{
        Various approaches to updating the covariance matrix $\Sigma$ of a tangent-space Gaussian distribution with mean $\mu$ to a mini-batch of samples.\@ We visualize the Gaussian distribution as well as the mini-batch in world space (top row) and in the $\mu$-centered tangent space (bottom row).
        \textbf{(a)} The state of the Gaussian distribution prior to the EM step.
        \textbf{(b)} When $\Sigma$ is na\"{\i}vely updated using the regular EM equations~(\ref{eq:em-ss-computation})~and~(\ref{eq:em-params-from-ss}), it admits an incorrect shape due to the distortion of mapping from the tangent space centered around the old mean $\mu$ to the one centered around the new mean $\hat{\mu}$.
        \textbf{(c)} Updating $\Sigma$ in the tangent space centered around the new mean $\hat{\mu}$ as per \autoref{eq:tangent-space-sigma-update} results in an accurate fit but is computationally costly, because the mini-batch must be transformed into the $\hat{\mu}$-centered tangent space first.
        \textbf{(d)} Updating $\Sigma$ using the regular EM equations (same as $\textbf{(b)}$) and subsequently transforming it into the $\hat{\mu}$-centered tangent space using our first-order approximation from \autoref{eq:product-distribution} results in an accurate fit with much lower cost.
    }\label{fig:covariance-foreign}
\end{figure*}

\begin{figure}
    \vspace{-6mm}
    \centering
    \begin{tikzpicture}
        \node[inner sep=0pt] (prod_both) at (0\linewidth,0.4\linewidth)
        {\includegraphics[width=0.30\linewidth]{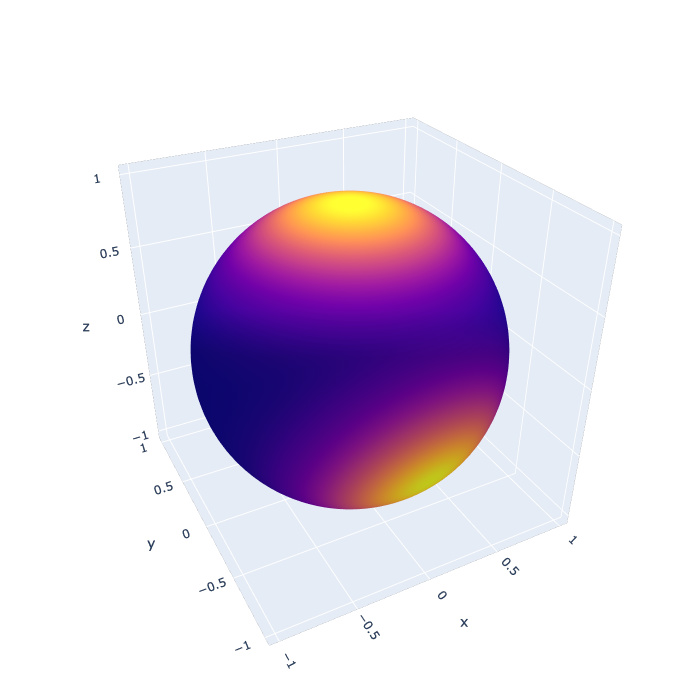}};
        \node[inner sep=0pt] (prod) at (0.4\linewidth,0.4\linewidth)
        {\includegraphics[width=0.30\linewidth]{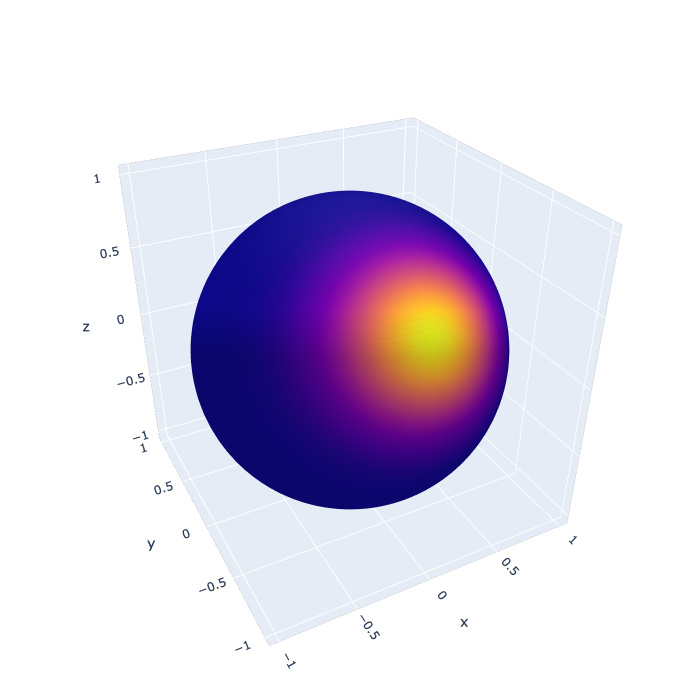}};

        \node[inner sep=0pt] (prod_both_t) at (0\linewidth,0\linewidth)
        {\includegraphics[width=0.30\linewidth]{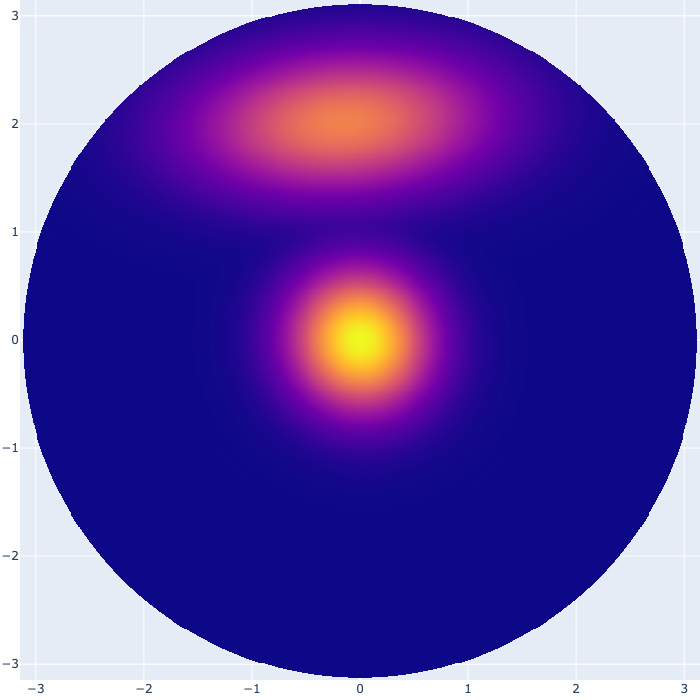}};
        \node[inner sep=0pt] (prod_t) at (0.4\linewidth,0\linewidth)
        {\includegraphics[width=0.30\linewidth]{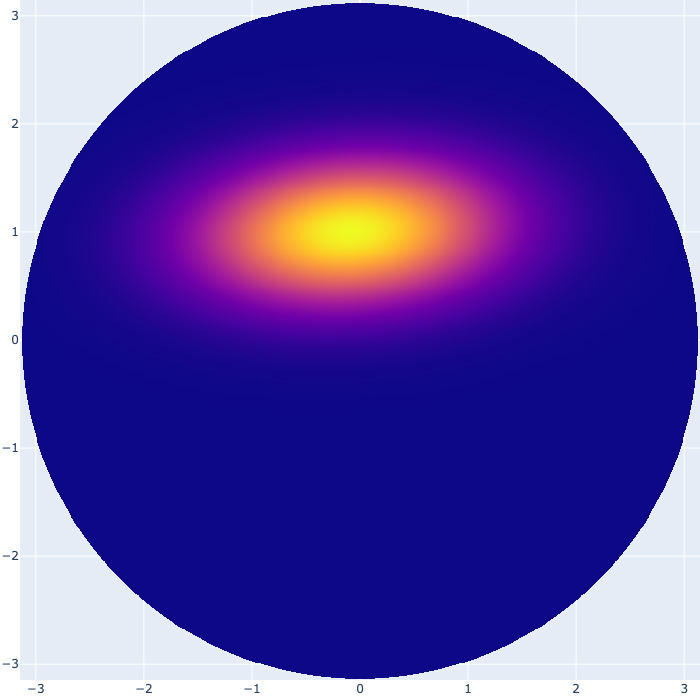}};

        \draw[->,thick] (prod_both.south) -- (prod_both_t.north)
            node[right,midway] {$\toTangent_{\mu_1}$};
        \draw[->,thick] (prod_both_t.east) -- (prod_t.west)
            node[midway,above] {mult.};
        \draw[->,thick] (prod_t.north) -- (prod.south)
        node[right,midway] {$\toWorld_{\mu_1}$};
    \end{tikzpicture}
    \caption{
        The product of two tangent-space Gaussians that are parameterized by $(\mu_1, \Sigma_1)$ and $(\mu_2, \Sigma_2)$ is computed in the $\mu_1$-centered tangent space according to \autoref{eq:product-distribution}.
        Then, samples for path guiding can be drawn and mapped to world-space by the $\toWorld_{\mu_1}$-map.
    }\vspace{-3mm}
    \label{fig:product}
\end{figure}

\subsection{Product Sampling from Mixed-Orientation Gaussians}%
\label{sec:product-sampling}

After conditioning $\PdfRadiance(\diri \,|\, \pos)$ and $\PdfBSDF(\diri \,|\, \diro, \bsdfParams)$, the last step is computing their product distribution such that it can be importance sampled (for the details of how $\PdfBSDF$ is conditioned in practice, please see Section~\ref{ss:bsdf-fitting}).

To do so, we must first rotate both mixtures into the same coordinate frame, which is achieved by applying the appropriate rotation matrix to the mean vectors $\mu_k$ of either the incident-radiance or the BSDF distribution.
In the following, we will assume that such a rotation has already been performed and the two distributions are in the same directional coordinate frame.

To compute the product distribution of the radiance approximation $\PdfRadiance(\diri \,|\, \pos)$ and the BSDF approximation $\PdfBSDF(\diri \,|\, \diro, \bsdfParams)$, one has to compute the product between \emph{each pair} of mixture components.
Let us consider a single such pair, where one component is parameterized by $(\mu_1, \Sigma_1)$ and the other component is parameterized by $(\mu_2, \Sigma_2)$.

To compute the product of these two components, they must be expressed within the same tangent space.
Without loss of generality, we choose to parameterize the second component in the $\mu_1$-centered tangent space.

\paragraph*{Mapping covariances to other tangent spaces.}
In the $\mu_1$-centered tangent space, the second mean vector $\mu_2$ is easily expressed using the $\toWorld$-map as $\toWorld_{\mu_1}(\mu_2)$,
but the question of how $\Sigma_2$ should transform from one tangent space to the other is more difficult to answer.
The most na\"ive approach would be to simply leave it unchanged, which we found to be too inaccurate; see \autoref{fig:cov-naive}.
The correct approach, on the other hand, would be to re-fit the covariance matrix through EM in the new tangent space, which would be impractically slow to do for each sampling decision.

We aim for a middle ground by considering the $\R^2 \mapsto \R^2$ mapping from one tangent space to the other:
\begin{align}
    \nu_1 = \toTangent_{\mu_1} \toWorld_{\mu_2}( \nu_2 )
\end{align}
If this mapping was \emph{linear}, it could be represented by a $2\!\times\!2$ matrix $J$ and the correct transformation of the covariance matrix could be computed efficiently as $J \Sigma_2 J^\intercal$.
However, the mapping is not actually linear, so we propose setting $J$ to a \emph{local linear approximation}---the Jacobian matrix---of $\toTangent_{\mu_1} \toWorld_{\mu_2}$.
We demonstrate the accuracy of this approach in \autoref{fig:cov-first-order},
where we compute $J$ as $J := J_{\toTangent_{\mu_1}}\!\!(\toWorld_{\mu_2}(0))\,J_{\toWorld_{\mu_2}}\!\!(0)$ with $J_{\toTangent}$ and $J_{\toWorld}$ being defined in \autoref{app:jacobians}.

Thus, the product distribution of a pair of components defined by $(\mu_1, \Sigma_1)$ and $(\mu_2, \Sigma_2)$ can be approximated in the $\mu_1$-centered tangent space as the product of the following two Gaussian distributions:
\renewcommand{\fboxsep}{4pt}
\begin{empheq}[box=\fbox]{equation}
\begin{split}
    \mathcal{N}\!\big(0, \Sigma_1\big) \otimes \mathcal{N}\!\big(\toTangent_{\mu_1}(\mu_2), J  \Sigma_2  J^\intercal \big) \,,
    \label{eq:product-distribution}
\end{split}
\end{empheq}
\renewcommand{\fboxsep}{0pt}%
\autoref{fig:product} illustrates this procedure.
Here, it is worth pointing out that the approximation error of our proposed covariance transformation $J  \Sigma_2  J^\intercal$ only applies to the \emph{second} mixture component.
Thus, in practice, however small the approximation error may be, one could choose the second mixture component to be the one where the error is most tolerable.
Alternatively, one might consider the computational cost of computing $J \Sigma_2  J^\intercal$. In our case, due to our usage of SIMD vectorization, it is benefitial to choose the second mixture to be the one with fewer components, enabling us to vectorize over the components of the first mixture.
We obtained the best overall efficiency by setting the first mixture to the incident radiance and the second mixture to the BSDF.\@

\paragraph*{Efficient tangent-space covariance update.}
The first-order approximation that enables accurate product sampling can also be used improve performance:
in \autoref{sec:map-em-with-tangent-spaces}, we state that the updated covariance matrix $\hat{\Sigma}$ must be computed in the tangent space centered around the newly computed mean $\hat{\mu}$ as opposed to the previous mean $\mu$; see \autoref{eq:tangent-space-sigma-update}.
This computation is expensive, because every data point in the mini batch as well as its corresponding probability density must be mapped into the $\hat{\mu}$-centered tangent space.
We side-step this expensive computation by computing $\hat{\Sigma}$ in the original $\mu$-centered tangent space and then approximately transforming it into the $\hat{\mu}$-centered tangent space using the same first-order approximation that we use in \autoref{eq:product-distribution} for product sampling.
In \autoref{fig:covariance-foreign}, we visually demonstrate the accuracy of this approach.

\section{Implementation in a Renderer}%
\label{sec:implementation}

As mentioned in \autoref{sec:methodology}, our algorithm requires us to train both a BSDF model $\PdfBSDF(\diri, \diro, \bsdfParams)$ as well a model of the incident radiance $\PdfRadiance(\diri, \pos)$.
In this section, we detail these procedures.

\subsection{BSDF Learning}%
\label{ss:bsdf-fitting}

Our approach to BSDF learning allows us, in theory, to fit a single $n$D mixture model to each type of BSDF that the renderer supports, where $n$ is the number of BSDF parameters.
For each sampling decision we can then specialize the previously learned general BSDF models by conditioning them on-the-fly on their parameters $\bsdfParams(\pos)$ at the shading location $\pos$.
This way, we avoid a scene-specific pre-computation of BSDFs, as well as opaquely handle both spatially-uniform and spatially-varying BSDFs.

In practice, to keep the number of dimensions manageable, we limit ourselves to isotropic BSDFs.
Our model has not been tested on anisotropic BSDFs.
It might be necessary to increase the number of Gaussian components to represent such models, which would be computationally prohibitive for our current approach.
One could use a $k$D-tree over the BSDF parameter space, similar to the one we use over the spatial dimensions of incident radiance,
or an alternative pruning framework, such as that of Herholz et al.~\cite{Herholz:2018}; see the discussion in \autoref{sec:discussion} for more details.

\paragraph*{BSDF input sampling.}
A BSDF mixture model is fitted by repeatedly sampling batches of BSDF inputs and then applying mini-batch EM to the mixture model.
Random BSDF parameters $\bsdfParams$ as well as the spherical coordinates of the outgoing direction $\diro$ are sampled uniformly.
Subsequently, $\diri$ is sampled using the BSDF's built-in importance sampling routine and mini-batch EM is applied.

\paragraph*{Pruning before product sampling.}
Since product sampling requires computing the product distribution between \emph{each pair} of mixture components within the BSDF and the incident-radiance model, the computational cost of product sampling can grow quadratically in the number of overall mixture components.
To suppress this quadratic growth, we employ a simple but effective strategy: after conditioning the BSDF model on its parameters, we perform product sampling using \emph{only} the two resulting BSDF mixture components with \emph{largest} magnitude.

\subsection{Online Learning of the Radiance During Path Tracing} \label{ss:radiance-fitting}

Path tracing of an $N$-spp image is performed in a fixed number of iterations that compute ${\nSppPerIteration}$ spp each.
The $i$-th iteration is importance sampled (i.e.\ path guided) using the product of the current state $\PdfRadiance^{i}$ of the incident-radiance model and the pre-computed BSDF distribution $\PdfBSDF$, each conditioned on-the-fly on the local shading parameters (see Sections \ref{sec:s-tree} and \ref{sec:product-sampling}).
The purpose of dividing path tracing into iterations is to facilitate periodic refinement of the incident-radiance model: after each iteration, the $k$D-tree is adapted to the path distribution and a subsequent mini-batch EM step is performed to produce a better distribution $\PdfRadiance^{i+1}$ for importance sampling in the next iteration.

\paragraph*{Spatial component of the radiance distribution.}
Our $5$D SDMMs should learn $\PdfRadiance(\diri, \pos) = \PdfRadiance(\diri \,|\, \pos) \PdfRadiance(\pos)$. However, during rendering, we only have access to $\PdfRadiance(\diri \,|\, \pos)$ in closed form.
The spatial distribution of samples, $\PdfRadiance(\pos)$, is defined by the ray-tracing procedure and is not known in a general setting.
While it would be possible to approximate the spatial distribution using a $k$-nearest-neighbor estimate, and then use that estimate to approximate the correct Monte Carlo weights in Eq. \ref{eq:em-ss-computation}~\cite{Dodik2020PathGU}, we have found the computational overhead of this approach to overshadow any possible gains due the quality of the learned distribution.
Instead, in our work, we assume a uniform distribution within each kD-tree leaf node.
In practice, this assumption becomes increasingly correct as the size of the kD-tree leaf nodes decreases during training.

\paragraph*{Refinement of the spatial $k$D-tree.}
During each iteration, we collect the encountered path vertices such that at the end of the iteration, each $k$D-tree leaf contains a list of all vertices that were located within it.
If the \emph{number} of path vertices $M$ in any leaf is larger than a threshold $c$, that leaf is subdivided along the axis with the largest variation of vertex positions, where the subdivision plane is located at the mean position along that axis~\cite{Herholz:2019}.
After subdivision, the vertices are redistributed into the two new leaves, which are then further subdivided recursively until they have fewer than $c$ vertices.
The mixture model from the parent node is \emph{copied} into the newly created leaves---the subsequent EM procedure will adapt each copy to the different light distributions in each leaf.
We empirically found ${c = 16000}$ to yield good results, which is of similar magnitude as the subdivision constant used in PPG~\cite{mueller2017practical}.

\paragraph*{Mini-batch EM fitting.}
Before the next iteration starts, we select all leaves that accumulated ${M \geq 16}$ vertices---for numeric stability---and independently apply a \emph{single} mini-batch EM step to each of them, where the mini-batch consists of all $M$ vertices contained in the respective leaf node. The vertices are subsequently cleared.
The EM step follows the procedure outlined in \autoref{sec:map-em-with-tangent-spaces}, where each vertex's incident-radiance estimate corresponds to its weights $w$, and its spatial and directional coordinates correspond to Euclidean and tangent-space dimensions, respectively.

Note, that the EM step can be computed in parallel across all leaf nodes, because the mixtures are independent from each other.

\paragraph*{Early stopping of training.}
Our incident-radiance mixture model converges to a good guiding distribution relatively early in the rendering process.
Thus, we stop the mini-batch EM training after the first $1/4$th of the rendering budget is exhausted to avoid its cost.
During the last $3/4$ths of the rendering budget, our mixture model is no longer trained and merely used for guiding.

\subsection{Robust EM Optimization of the Incident-Radiance PDF} \label{ss:robust-em}

While the EM algorithm is theoretically elegant and converges quickly, it is prone to get stuck in undesirable local minima, making it unstable in our online-learning setting.
The instability becomes worse when operating on noisy Monte Carlo weights, which are common in rendering when estimating incident radiance (e.g.\ ``fireflies'').
We therefore found it paramount to employ the following array of regularization techniques to robustify the EM training of our incident radiance approximation.\@

\paragraph*{Initialization.}
We aim at initializing the Gaussian mixture such that it evenly covers the $5$D domain (no ``clumping'') to prevent EM from converging to a local optimum that misses a mode of $\inRadiance$.
To this end, and to avoid lock contention,
prior to rendering, the $k$D-tree is pre-subdivided $3$ times at the center of each axis, resulting in a regular ${8\!\times\!8\!\times\!8}$ tesselation of the scene's bounding box.
After each render iteration, the we split the $k$D-tree and distribute the collected samples to the newly created leaf nodes. If a leaf node is uninitialized and contains at least $16$ path vertices, we initialize a mixture model containing $\nGaussians$ Gaussians inside of that leaf node. If a leaf node contains fewer than $16$ samples, the initialization of the corresponding mixture model is postponed until $16$ samples were received to ensure stability in noisy scenes.

After $16$ vertices were received, initialization consists of the following three steps.
First, a slightly modified \texttt{k-means++}~\cite{kmeans} scheme is applied to the vertices to select $2$ spatial coordinates $\pos_1, \pos_2$ that lie on a geometric surface, have a minimum distance from each other (if possible), and that roughly follow the spatial distribution of radiance; see \autoref{app:k-means} for details.
Then, at each one of the 2 chosen positions, $8$ Gaussians are initialized whose directional mean components $\dir_1, \ldots, \dir_8$ cover the sphere roughly evenly.
Formally, the set of initial mean vectors is $\{ \pos_1, \pos_2 \} \times \{ \dir_1, \ldots, \dir_8 \}$.

Third, the covariance matrix at each mean vector $(\pos,\dir)$ is initialized to satisfy the following desired properties. It should be
\begin{itemize}
    \item isotropic in the tangent plane of the surface at $\pos$ and in the $\dir$-centered directional tangent space, and
    \item flat along the surface normal at $\pos$.
\end{itemize}
To this end, in the tangent plane, the covariance matrix's radius is set to be proportional to the size of the leaf node containing it, and inversely proportional to the number of distinct spatial coordinates in the mixture (in our case $2$).
In contrast, we make the covariance matrix as thin as possible along the direction of the surface normal, to avoid mixing samples from close-by surfaces facing each other.
In the remaining $\dir$-centered tangent space dimensions, the covariance matrix's radius is inversely proportional to the number of distinct directional coordinates in the mixture model, i.e.\ $8$.
See \autoref{app:cov-init} for the mathematical details of this procedure.

\paragraph*{Averaging of sufficient statistics across minibatches.}
In \autoref{sec:map-em-with-tangent-spaces}, we described the computation of a per-mixture-component triplet of sufficient statistics $(S^{(0)}, S^{(1)}, S^{(2)})$ as sample averages over mini batches.
In practice, such sample averages are often too noisy to be directly useful for EM optimization.
Thus, similar to previous work~\cite{Capp__2009, cappe:hal-00532968, Vorba:2014:OnlineLearningPMMinLTS}, we additionally average the sufficient statistics \emph{across} mini batches using the Robbins-Monroe algorithm~\cite{robbinsmonro}, which ensures that EM converges under mild assumptions\footnote{The assumptions include a uniform bound on $S^{(i)}$, which is not necessarily met in rendering. Nonetheless, we observe stable behavior in practice.}.
Let $\hat{S}^{(i)}_{k, j}$ be the averaged sufficient statistics of the $k$\textsuperscript{th} mixture component after processing the $j$-th mini batch; they are computed as
\begin{align}
    \hat{S}^{(i)}_{k, j} = (1 - \ssEmaDecay_j) \cdot \hat{S}^{(i)}_{k, j-1} + \ssEmaDecay_j \cdot S^{(i)}_k \,,
    \label{eq:robbins-monroe}
\end{align}
where $\ssEmaDecay_j$ controls the strength of averaging.
While any sequence $\ssEmaDecay_j$ that satisfies $\sum_j^\infty \ssEmaDecay_j = \infty \wedge \sum_j^\infty \ssEmaDecay_j^2 < \infty$ makes EM converge in theory, we adopt the sequence $\ssEmaDecay_j = (\beta j + 1)^{-\alpha}$ inspired by common practice~\cite{Capp__2009, cappe:hal-00532968, Vorba:2014:OnlineLearningPMMinLTS,Nguyen_2020}.
We set ${\alpha = 0.5}$ as done in previous work, ${\beta = 0.1}$ to achieve a long tail (averaging over many iterations to reduce noise). 

Special care must be taken when averaging the second-moment sufficient statistics in tangent spaces.
The second-moment statistic of the current mini batch $S^{(2)}$ is defined in the tangent space centered about the \emph{current} mean vector $\mu_j$, whereas the previously computed statistic $\hat{S}^{(2)}_{j-1}$ is defined in the tangent space centered about the \emph{previous} mean-vector $\mu_{j-1}$.
Thus, prior to averaging, $\hat{S}^{(i)}_{j-1}$ must first be transformed into the $\mu_j$-centered tangent space.
To this end, we use the same first-order approximation that we use in \autoref{eq:product-distribution}, resulting in the following modified update rule for the tangent-space components of the second-moment sufficient statistics:
\begin{align}
    \hat{S}^{(2)}_j = \ssEmaDecay_j \cdot J \hat{S}^{(2)}_{j-1} J^\intercal + (1 - \ssEmaDecay_j) \cdot S^{(2)} \,.
    \label{eq:robbins-monroe-2nd}
\end{align}
In the above equation, $J$ is the Jacobian matrix of $\toTangent_{\mu_{j-1}} \toWorld_{\mu_j}$ at $0$.

Lastly, same as Vorba et al.~\cite{Vorba:2014:OnlineLearningPMMinLTS}, we also apply the Robbins-Monroe averaging scheme (with the same sequence $\ssEmaDecay_j$) to the approximate normalization factor that is used in the M-step.

\paragraph*{Maximum a-posteriori EM.}
To learn incident radiance, we use maximum a-posteriori (MAP) EM, which is an extension of the maximum-likelihood EM algorithm that we outlined in \autoref{sec:map-em-with-tangent-spaces}.
Unlike maximum-likelihood EM, MAP EM takes ``prior'' information about the mixture parameters $(\mixtureWeight_k, \mu_k, \Sigma_k)$ into account.
Such prior information regularizes the optimization by biasing EM towards solutions that resemble a chosen prior distribution.
Following Gauvain and Lee~\cite{Gauvain1994MaximumAP}, we choose a prior Dirichlet distribution over the weights $\mixtureWeight_k$ and a prior \ADD{inverse} Wishart distribution over the covariances $\Sigma_k$ (no prior is used over $\mu_k$).
This formulation results in a closed-form MAP EM optimization that matches maximum-likelihood EM, except for the following modifications to the tangent-space update formulae of the mixture weights $\hat{\mixtureWeight}_k$ and the covariance matrices $\hat{\Sigma}_k$:
\begin{align}
    \hat{\mixtureWeight}_k = \frac{\PriorScale_j\weightPrior + S_k^{(0)}}{\nGaussiansVar\PriorScale_j\weightPrior + \sum_i^\nGaussiansVar \, S_i^{(0)}} \,, \,\,\,\,\,
    \hat{\Sigma}_k = \frac{\PriorScale_j\covPrior + S_k^{(2)}}{\PriorScale_j\denomPrior + S_k^{(0)}} \,.
\end{align}
The scalars $\weightPrior$, $\denomPrior$ and the matrix $\covPrior$ parameterize the prior distributions and $\PriorScale_j$ progressively downscales the strength of the priors as the training step $j$ increases.
This downscaling is a consequence of adding more samples to a Bayesian estimator, and in practice removes the bias introduced by the prior distributions when sufficiently many data points have been observed for a robust maximum likelihood estimate.

We choose the prior parameters
\begin{align}
   \weightPrior = 1 / \nGaussiansVar \,, \,\,\,\,\, \denomPrior = 5 / \nGaussiansVar \,, \,\,\,\,\, \covPrior = \denomPrior \diag{(0.1, 0.1, 1, 1, 1)} \times 10^{-4} \,,
\end{align}
where the first two entries on the diagonal of $\covPrior$ are responsible for regularizing the directional tangent-space dimensions $\nu$ and the remaining entries are responsible for the spatial dimensions $\pos$---hence their different scale.

The relatively large Dirichlet prior $\weightPrior$ ensures that each mixture component is weighted roughly equally in the early stages of the optimization, whereas the relatively large Wishart prior parameter $\denomPrior$ ensures that the covariance matrix does not fluctuate too strongly due to initially chosen mixture weights.
Since $S_i^{(0)}$ is roughly inversely proportional to the number of mixture components $K$, we include a division by $K$ in our definitions of $\weightPrior$ and $\denomPrior$.
The effect is that the relative prior strength per mixture component is constant.

Our choice of a small Wishart prior parameter $\covPrior$ serves the purpose of preventing the covariance matrices from collapsing when few samples are available, e.g.\ onto a firefly.
Other than that (e.g. as soon as $S_k^{(0)}$ is sufficiently large), the small value of $\covPrior$ does not noticeably restrict the shape of the learned covariance matrices.
Thus, accurately tailoring the spatial components of $\covPrior$ to the scene scale is usually not necessary.
Nonetheless, we normalize $\pos$ to span the range ${[0, 1]}^3$ within the scene's bounding box, and make the initialization of the covariance matrices proportional to the size of the kD-tree leaf nodes containing them.
In extreme cases, such as an expansive outdoor environment, it could possibly be necessary to also additionally scale the spatial components of $\covPrior$ by the size of the corresponding kD-tree leaf node.
We experimented with such an approach, but had difficulties in making it work well across all our scenes due to numerical instabilities.

\begin{figure*}
    \captionof{table}{\label{tab:results}
        \ADD{Comparison of our method (SDMM radiance \& product) with practical path guiding (PPG)~\cite{mueller19guiding} and parallax-aware von Mises-Fisher mixtures (VMM radiance)~\cite{ruppert2020robust} using the respective authors' implementations.
        We rendered each scene with \num{1024} spp, where the training and rendering stages of each method (except for PPG) were allotted \num{252} and \num{772} spp.
        We report mean absolute percentage error (MAPE), render time, and the speedup vs.\ PPG at reaching the lowest common MAPE value.
        For Ruppert et al., we omit the speedup due to their separation of training and rendering. Instead, we report equal-time results in \autoref{fig:vmm}.
        The best entries are highlighted in \textbf{bold} letters.
        Our radiance-based method (SDMM radiance) achieves lower error than PPG and on average similar error as radiance-based VMMs, but at the cost of increased render time.
        Incorporating the product (SDMM product) reduces MAPE further while again increasing the render time.}
    }
    \vspace{-3.5mm}
\small
\rowcolors{2}{white}{gray!8}
\setlength{\tabcolsep}{16.1pt}
\begin{tabular*}{\textwidth}{rcccc}
  \toprule
  \rowcolor{white}
  {} & \cite{mueller19guiding} & \cite{ruppert2020robust} & \multicolumn{2}{c}{Ours} \\
  \cmidrule(lr){2-2}
  \cmidrule(lr){3-3}
  \cmidrule(lr){4-5}
  \rowcolor{white}
  {} & PPG & VMM radiance & SDMM radiance & SDMM product \\
  \midrule
  \small{\Bathroom{}} & \small{\makebox[0.75cm]{0.192} \, \makebox[0.75cm]{\textcolor{gray}{4.2m}} \, \makebox[0.75cm]{\textbf{(1.0$\times$)}}} & \small{\makebox[0.75cm]{0.148} \, \makebox[0.75cm]{\textcolor{gray}{3.1m}}} & \small{\makebox[0.75cm]{0.161} \, \makebox[0.75cm]{\textcolor{gray}{6.5m}} \, \makebox[0.75cm]{(0.9$\times$)}} & \small{\makebox[0.75cm]{\textbf{0.115}} \, \makebox[0.75cm]{\textcolor{gray}{12m}} \, \makebox[0.75cm]{(0.9$\times$)}} \\
  \small{\Bedroom{}} & \small{\makebox[0.75cm]{0.060} \, \makebox[0.75cm]{\textcolor{gray}{3.2m}} \, \makebox[0.75cm]{(1.0$\times$)}} & \small{\makebox[0.75cm]{0.062} \, \makebox[0.75cm]{\textcolor{gray}{2.7m}}} & \small{\makebox[0.75cm]{0.054} \, \makebox[0.75cm]{\textcolor{gray}{4.5m}} \, \makebox[0.75cm]{(0.9$\times$)}} & \small{\makebox[0.75cm]{\textbf{0.051}} \, \makebox[0.75cm]{\textcolor{gray}{7.1m}} \, \makebox[0.75cm]{(0.7$\times$)}} \\
  \small{\Bookshelf{}} & \small{\makebox[0.75cm]{0.132} \, \makebox[0.75cm]{\textcolor{gray}{4.6m}} \, \makebox[0.75cm]{(1.0$\times$)}} & \small{\makebox[0.75cm]{0.105} \, \makebox[0.75cm]{\textcolor{gray}{3.4m}}} & \small{\makebox[0.75cm]{0.111} \, \makebox[0.75cm]{\textcolor{gray}{6.6m}} \, \makebox[0.75cm]{\textbf{(1.0$\times$)}}} & \small{\makebox[0.75cm]{\textbf{0.085}} \, \makebox[0.75cm]{\textcolor{gray}{11m}} \, \makebox[0.75cm]{(1.0$\times$)}} \\
  \small{\Bottle{}} & \small{\makebox[0.75cm]{0.274} \, \makebox[0.75cm]{\textcolor{gray}{3.3m}} \, \makebox[0.75cm]{(1.0$\times$)}} & \small{\makebox[0.75cm]{\textbf{0.152}} \, \makebox[0.75cm]{\textcolor{gray}{2.5m}}} & \small{\makebox[0.75cm]{0.205} \, \makebox[0.75cm]{\textcolor{gray}{4.3m}} \, \makebox[0.75cm]{\textbf{(1.8$\times$)}}} & \small{\makebox[0.75cm]{0.171} \, \makebox[0.75cm]{\textcolor{gray}{9.5m}} \, \makebox[0.75cm]{(1.3$\times$)}} \\
  \small{\CornellBox{}} & \small{\makebox[0.75cm]{0.069} \, \makebox[0.75cm]{\textcolor{gray}{1.1m}} \, \makebox[0.75cm]{(1.0$\times$)}} & \small{\makebox[0.75cm]{0.044} \, \makebox[0.75cm]{\textcolor{gray}{43s}}} & \small{\makebox[0.75cm]{0.026} \, \makebox[0.75cm]{\textcolor{gray}{1.6m}} \, \makebox[0.75cm]{\textbf{(3.2$\times$)}}} & \small{\makebox[0.75cm]{\textbf{0.020}} \, \makebox[0.75cm]{\textcolor{gray}{2.2m}} \, \makebox[0.75cm]{(3.2$\times$)}} \\
  \small{\GlossyKitchen{}} & \small{\makebox[0.75cm]{0.316} \, \makebox[0.75cm]{\textcolor{gray}{2.8m}} \, \makebox[0.75cm]{(1.0$\times$)}} & \small{\makebox[0.75cm]{0.250} \, \makebox[0.75cm]{\textcolor{gray}{1.9m}}} & \small{\makebox[0.75cm]{0.205} \, \makebox[0.75cm]{\textcolor{gray}{4.4m}} \, \makebox[0.75cm]{\textbf{(1.7$\times$)}}} & \small{\makebox[0.75cm]{\textbf{0.151}} \, \makebox[0.75cm]{\textcolor{gray}{11m}} \, \makebox[0.75cm]{(1.3$\times$)}} \\
  \small{\Necklace{}} & \small{\makebox[0.75cm]{0.187} \, \makebox[0.75cm]{\textcolor{gray}{1.1m}} \, \makebox[0.75cm]{\textbf{(1.0$\times$)}}} & \small{\makebox[0.75cm]{\textbf{0.127}} \, \makebox[0.75cm]{\textcolor{gray}{1.1m}}} & \small{\makebox[0.75cm]{0.181} \, \makebox[0.75cm]{\textcolor{gray}{1.5m}} \, \makebox[0.75cm]{(0.8$\times$)}} & \small{\makebox[0.75cm]{0.157} \, \makebox[0.75cm]{\textcolor{gray}{2.7m}} \, \makebox[0.75cm]{(0.8$\times$)}} \\
  \small{\SwimmingPool{}} & \small{\makebox[0.75cm]{0.076} \, \makebox[0.75cm]{\textcolor{gray}{2.1m}} \, \makebox[0.75cm]{\textbf{(1.0$\times$)}}} & \small{\makebox[0.75cm]{0.079} \, \makebox[0.75cm]{\textcolor{gray}{2.0m}}} & \small{\makebox[0.75cm]{0.068} \, \makebox[0.75cm]{\textcolor{gray}{2.9m}} \, \makebox[0.75cm]{(0.9$\times$)}} & \small{\makebox[0.75cm]{\textbf{0.058}} \, \makebox[0.75cm]{\textcolor{gray}{4.0m}} \, \makebox[0.75cm]{(1.0$\times$)}} \\
  \small{\Torus{}} & \small{\makebox[0.75cm]{0.078} \, \makebox[0.75cm]{\textcolor{gray}{1.1m}} \, \makebox[0.75cm]{\textbf{(1.0$\times$)}}} & \small{\makebox[0.75cm]{0.085} \, \makebox[0.75cm]{\textcolor{gray}{1.1m}}} & \small{\makebox[0.75cm]{0.073} \, \makebox[0.75cm]{\textcolor{gray}{1.5m}} \, \makebox[0.75cm]{(0.9$\times$)}} & \small{\makebox[0.75cm]{\textbf{0.072}} \, \makebox[0.75cm]{\textcolor{gray}{2.0m}} \, \makebox[0.75cm]{(0.8$\times$)}} \\
  \small{\VeachDoor{}} & \small{\makebox[0.75cm]{0.221} \, \makebox[0.75cm]{\textcolor{gray}{2.1m}} \, \makebox[0.75cm]{\textbf{(1.0$\times$)}}} & \small{\makebox[0.75cm]{0.286} \, \makebox[0.75cm]{\textcolor{gray}{1.5m}}} & \small{\makebox[0.75cm]{0.188} \, \makebox[0.75cm]{\textcolor{gray}{3.6m}} \, \makebox[0.75cm]{(0.8$\times$)}} & \small{\makebox[0.75cm]{\textbf{0.122}} \, \makebox[0.75cm]{\textcolor{gray}{8.3m}} \, \makebox[0.75cm]{(0.8$\times$)}} \\
  \small{\WaterCaustic{}} & \small{\makebox[0.75cm]{0.549} \, \makebox[0.75cm]{\textcolor{gray}{1.8m}} \, \makebox[0.75cm]{(1.0$\times$)}} & \small{\makebox[0.75cm]{0.639} \, \makebox[0.75cm]{\textcolor{gray}{1.9m}}} & \small{\makebox[0.75cm]{0.379} \, \makebox[0.75cm]{\textcolor{gray}{2.6m}} \, \makebox[0.75cm]{\textbf{(2.8$\times$)}}} & \small{\makebox[0.75cm]{\textbf{0.370}} \, \makebox[0.75cm]{\textcolor{gray}{3.8m}} \, \makebox[0.75cm]{(2.4$\times$)}} \\
  \small{\GlossyCornellBox{}} & \small{\makebox[0.75cm]{0.560} \, \makebox[0.75cm]{\textcolor{gray}{1.3m}} \, \makebox[0.75cm]{(1.0$\times$)}} & \small{\makebox[0.75cm]{0.080} \, \makebox[0.75cm]{\textcolor{gray}{1.1m}}} & \small{\makebox[0.75cm]{0.068} \, \makebox[0.75cm]{\textcolor{gray}{2.4m}} \, \makebox[0.75cm]{\textbf{(15.5$\times$)}}} & \small{\makebox[0.75cm]{\textbf{0.050}} \, \makebox[0.75cm]{\textcolor{gray}{5.2m}} \, \makebox[0.75cm]{(9.2$\times$)}} \\
  \bottomrule
\end{tabular*}

    \begin{subfigure}{\linewidth}
        \vspace{0mm}
        \hspace*{-1mm}\includegraphics[width=1.01\textwidth]{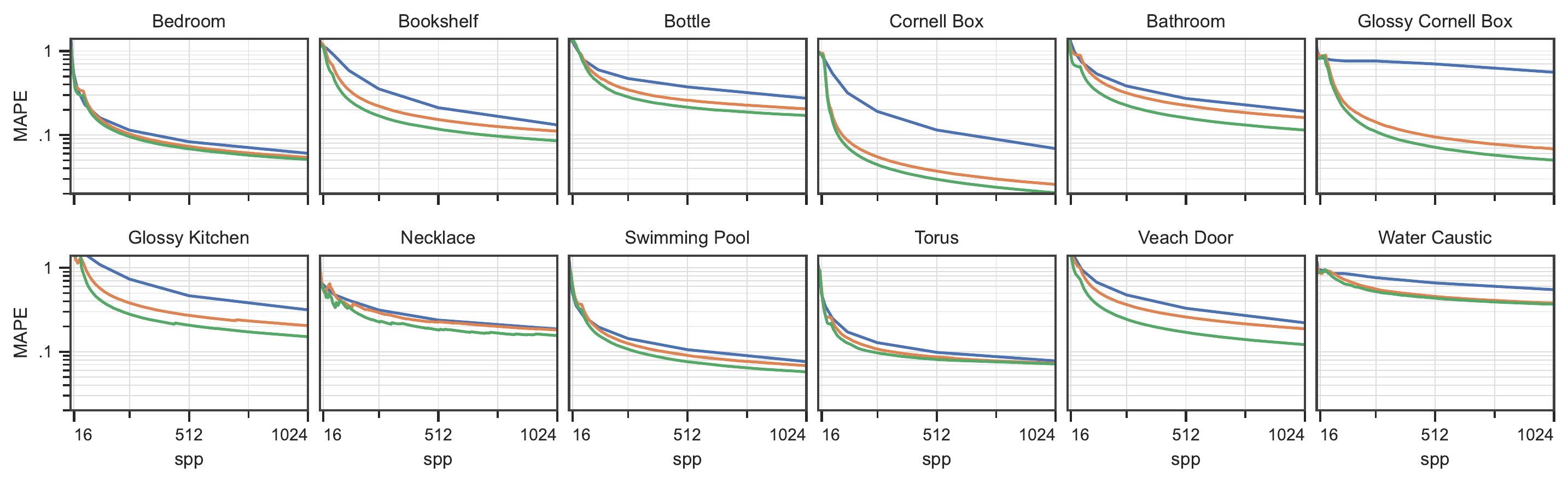}
        \vspace{-5.5mm}%
        \caption{MAPE vs.\ samples per pixel}
    \end{subfigure}
    \begin{subfigure}{\linewidth}
        \vspace{1mm}
        \hspace*{-1mm}\includegraphics[width=1.01\textwidth]{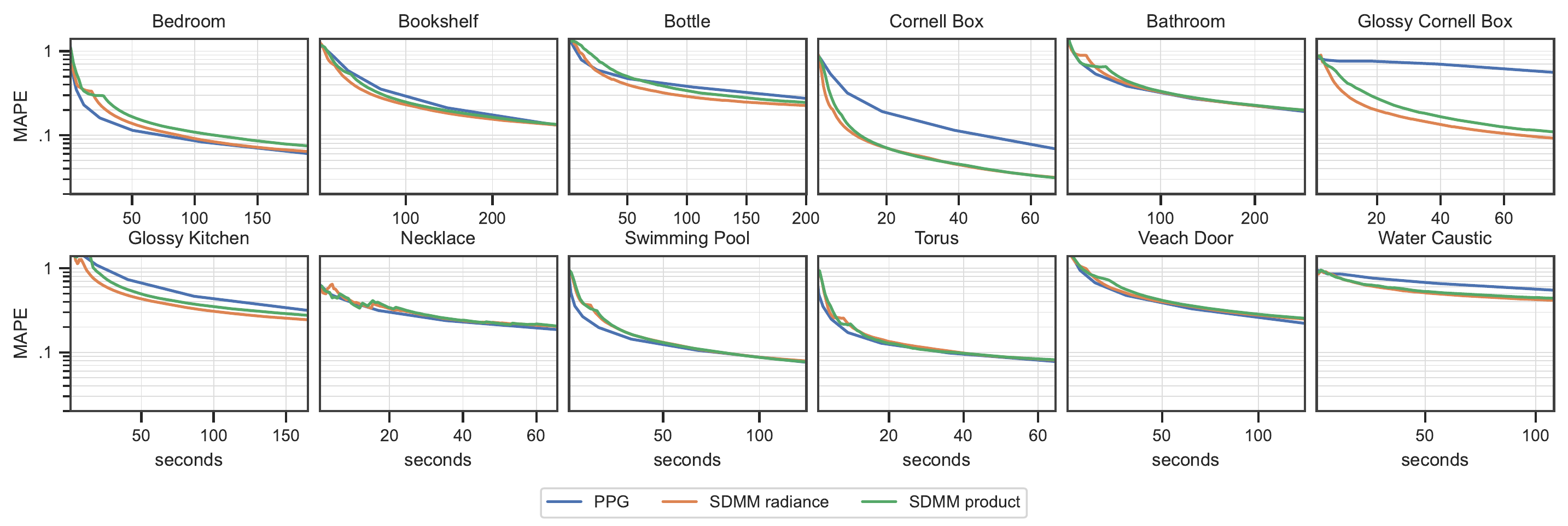}
        \vspace{-4.5mm}%
        \caption{MAPE vs.\ render time}
    \end{subfigure}
    \vspace{-1mm}
    \captionof{figure}{\label{fig:convergence}
        We analyze the convergence behavior of our radiance- and product-based guiding algorithms by plotting MAPE (a) against the number of samples per pixel and (b) against the render time.
        (a) At equal sample counts, our algorithms consistently outperform PPG.\@
        (b) However, at equal render time, we are mostly on par with PPG due to our larger computational overhead. In scenes with strong spatio-directional correlation in the incident radiance, our algorithm significantly outperforms PPG (\CornellBox{}, \Bookshelf{}, \WaterCaustic{}, \GlossyCornellBox{}, \GlossyKitchen{}).
    }
    \vspace{-2mm}
\end{figure*}

\begin{figure*}
\setlength{\fboxrule}{10pt}%
\setlength{\insetvsep}{20pt}%
\setlength{\tabcolsep}{-1pt}%
\renewcommand{\arraystretch}{1}%
\small%
\hspace*{3mm}%
\begin{tabular}{rcccccc}
  {} & {} & PPG & VMM radiance & SDMM radiance & SDMM product & Reference \\
    \setInset{A}{red}{140}{150}{61}{38}%
    \setInset{B}{orange}{483}{90}{61}{38}%
    \rotatebox{90}{\hspace{-1.85cm}\Bookshelf{}}\hspace{0.14cm} &
    \addBeautyCrop{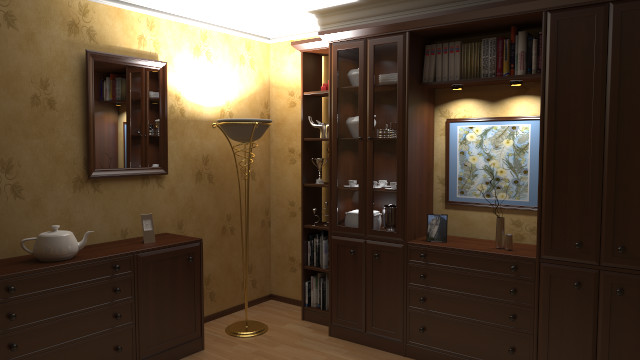}{0.29}{640}{360}{0}{0}{640}{360} &
    \addInsets{images/fig-results/bookshelf-PPG_u.jpg} &
    \addInsets{images/fig-results/bookshelf-VMM-radiance_u.jpg} &
    \addInsets{images/fig-results/bookshelf-SDMM-radiance_u.jpg} &
    \addInsets{images/fig-results/bookshelf-SDMM-product_u.jpg} &
    \addInsets{images/fig-references/bookshelf-reference.jpg} \\
  & \multicolumn{1}{r}{MAPE:} & 0.132 & 0.105 & 0.111 & \textbf{0.085} & \\
  & \multicolumn{1}{r}{render time:} & 4.6m & 3.4m & 6.6m & 11m & \\
  & \multicolumn{1}{r}{speedup vs.\ PPG at equal MAPE:} & 1.000 & \small{---} & \textbf{1.005} & 0.965 & \\
    \setInset{A}{red}{233}{133}{61}{38}%
    \setInset{B}{orange}{443}{225}{61}{38}%
    \rotatebox{90}{\hspace{-2.20cm}\GlossyKitchen{}}\hspace{0.14cm} &
    \addBeautyCrop{images/fig-references/glossy-kitchen-reference-lq.jpg}{0.29}{640}{360}{0}{0}{640}{360} &
    \addInsets{images/fig-results/glossy-kitchen-PPG_u.jpg} &
    \addInsets{images/fig-results/glossy-kitchen-VMM-radiance_u.jpg} &
    \addInsets{images/fig-results/glossy-kitchen-SDMM-radiance_u.jpg} &
    \addInsets{images/fig-results/glossy-kitchen-SDMM-product_u.jpg} &
    \addInsets{images/fig-references/glossy-kitchen-reference.jpg} \\
  & \multicolumn{1}{r}{MAPE:} & 0.316 & 0.250 & 0.205 & \textbf{0.151} & \\
  & \multicolumn{1}{r}{render time:} & 2.8m & 1.9m & 4.4m & 11m & \\
  & \multicolumn{1}{r}{speedup vs.\ PPG at equal MAPE:} & 1.000 & \small{---} & \textbf{1.731} & 1.303 & \\
    \setInset{A}{red}{25}{6}{61}{38}%
    \setInset{B}{orange}{250}{200}{61}{38}%
    \rotatebox{90}{\hspace{-2.15cm}\SwimmingPool{}}\hspace{0.14cm} &
    \addBeautyCrop{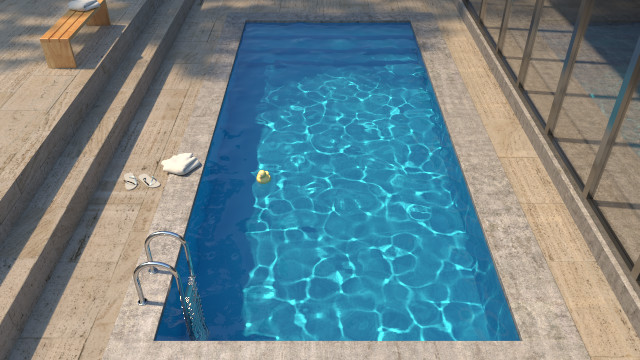}{0.29}{640}{360}{0}{0}{640}{360} &
    \addInsets{images/fig-results/pool-PPG_u.jpg} &
    \addInsets{images/fig-results/pool-VMM-radiance_u.jpg} &
    \addInsets{images/fig-results/pool-SDMM-radiance_u.jpg} &
    \addInsets{images/fig-results/pool-SDMM-product_u.jpg} &
    \addInsets{images/fig-references/pool-reference.jpg} \\
  & \multicolumn{1}{r}{MAPE:} & 0.076 & 0.079 & 0.068 & \textbf{0.058} & \\
  & \multicolumn{1}{r}{render time:} & 2.1m & 2.0m & 2.9m & 4.0m & \\
  & \multicolumn{1}{r}{speedup vs.\ PPG at equal MAPE:} & \textbf{1.000} & \small{---} & 0.917 & 0.981 & \\
    \setInset{A}{red}{158}{225}{61}{38}%
    \setInset{B}{orange}{378}{216}{61}{38}%
    \rotatebox{90}{\hspace{-2.0cm}\VeachDoor{}}\hspace{0.14cm} &
    \addBeautyCrop{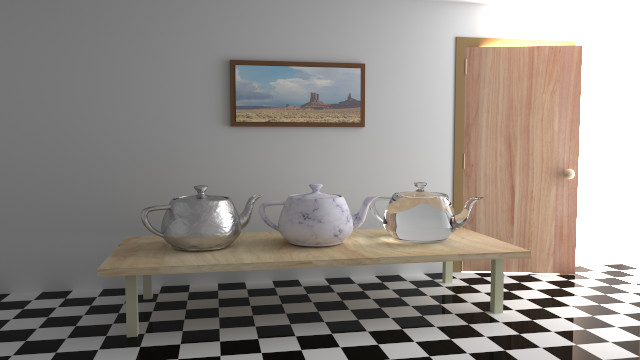}{0.29}{640}{360}{0}{0}{640}{360} &
    \addInsets{images/fig-results/veach-door-PPG_u.jpg} &
    \addInsets{images/fig-results/veach-door-VMM-radiance_u.jpg} &
    \addInsets{images/fig-results/veach-door-SDMM-radiance_u.jpg} &
    \addInsets{images/fig-results/veach-door-SDMM-product_u.jpg} &
    \addInsets{images/fig-references/veach-door-reference.jpg} \\
  & \multicolumn{1}{r}{MAPE:} & 0.221 & 0.286 & 0.188 & \textbf{0.122} & \\
  & \multicolumn{1}{r}{render time:} & 2.1m & 1.5m & 3.6m & 8.3m & \\
  & \multicolumn{1}{r}{speedup vs.\ PPG at equal MAPE:} & \textbf{1.000} & \small{---} & 0.795 & 0.789 & \\
    \setInset{A}{red}{400}{33}{61}{38}%
    \setInset{B}{orange}{333}{300}{61}{38}%
    \rotatebox{90}{\hspace{-2.2cm}\WaterCaustic{}}\hspace{0.14cm} &
    \addBeautyCrop{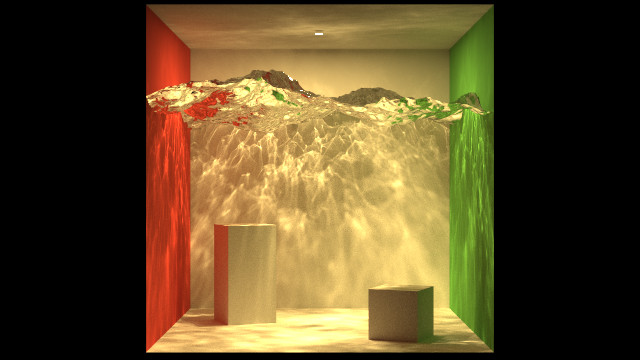}{0.29}{640}{360}{0}{0}{640}{360} &
    \addInsets{images/fig-results/water-caustic-PPG_u.jpg} &
    \addInsets{images/fig-results/water-caustic-VMM-radiance_u.jpg} &
    \addInsets{images/fig-results/water-caustic-SDMM-radiance_u.jpg} &
    \addInsets{images/fig-results/water-caustic-SDMM-product_u.jpg} &
    \addInsets{images/fig-references/water-caustic-reference.jpg} \\
  & \multicolumn{1}{r}{MAPE:} & 0.549 & 0.639 & 0.379 & \textbf{0.370} & \\
  & \multicolumn{1}{r}{render time:} & 1.8m & 1.9m & 2.6m & 3.8m & \\
  & \multicolumn{1}{r}{speedup vs.\ PPG at equal MAPE:} & 1.000 & \small{---} & \textbf{2.818} & 2.416 & \\
\end{tabular}

    \caption{\label{fig:images}
        \ADD{Visual comparison of the same experimental setup as in \autoreftable{tab:results}.
        For Ruppert et al., we omit the speedup due to their separation of training and rendering.
        Instead, we report equal-time results in \autoref{fig:vmm}.
        Our radiance-based method (SDMM radiance) consistently achieves lower error than PPG and on average similar error as radiance-based VMMs.
        Incorporating the product (SDMM product) reduces the error further.
        However, the computational overhead of our methods is larger than that of the other methods, leading to VMMs performing best at equal time.
        As expected, our methods perform better when the incident radiance exhibits significant spatio-directional correlation (\WaterCaustic{}, \Bookshelf{}, and \GlossyKitchen{}) and performs worse in the opposite case, e.g.\ environment lighting in the \SwimmingPool{}.}
    }
\end{figure*}


\section{Results}%
\label{sec:results}

We implemented our method within the Mitsuba renderer~\cite{Mitsuba}, making heavy use of SIMD vectorization via Enoki~\cite{Enoki} for computations pertaining to our mixture models.
Our reference implementation will be released publicly upon publication of this work.

All images in these section were generated at a resolution of $640\!\times\!360$ pixels, using \num{1024} samples per pixel, on an Intel Xeon W-2135 CPU with $12$ cores and $64$ GB RAM.
The reference images were rendered with standard path tracing using a very large sample count that was chosen for each scene such that the reference image has no visible left-over noise.
All comparisons have next-event estimation and Russian roulette disabled and the number of path vertices is bounded to $10$.

\begin{figure}
\begin{minipage}{\columnwidth}
    \captionof{table}{\label{tab:vmm}
        Comparison of radiance-based VMMs~\cite{ruppert2020robust} with our SDMMs at equal training and rendering time.
        For both methods, we report mean absolute percentage error (MAPE) as well as training + rendering time.
        Due to their much lower computational cost, VMMs produce lower error than SDMMs on most scenes.
    }
    \vspace{0mm}
    \small
\rowcolors{2}{white}{gray!8}
\setlength{\tabcolsep}{3.00pt}
\begin{tabular*}{\columnwidth}{rcc}
  \toprule
  \rowcolor{white}
  {} & Ruppert et al.~\cite{ruppert2020robust} & SDMM Radiance (Ours) \\
  \midrule
  \small{\Bathroom{}} & \small{\makebox[0.7cm]{\textbf{0.104}} \, \textcolor{gray}{\makebox[0.5cm]{1.8m} + \makebox[0.5cm]{4.7m}}} & \small{\makebox[0.7cm]{0.179} \, \textcolor{gray}{\makebox[0.5cm]{1.8m} + \makebox[0.5cm]{4.7m}}} \\
  \small{\Bedroom{}} & \small{\makebox[0.7cm]{\textbf{0.054}} \, \textcolor{gray}{\makebox[0.5cm]{1.4m} + \makebox[0.5cm]{3.0m}}} & \small{\makebox[0.7cm]{0.059} \, \textcolor{gray}{\makebox[0.5cm]{1.4m} + \makebox[0.5cm]{3.0m}}} \\
  \small{\Bookshelf{}} & \small{\makebox[0.7cm]{\textbf{0.080}} \, \textcolor{gray}{\makebox[0.5cm]{1.8m} + \makebox[0.5cm]{4.8m}}} & \small{\makebox[0.7cm]{0.121} \, \textcolor{gray}{\makebox[0.5cm]{1.8m} + \makebox[0.5cm]{4.8m}}} \\
  \small{\Bottle{}} & \small{\makebox[0.7cm]{\textbf{0.116}} \, \textcolor{gray}{\makebox[0.5cm]{1.1m} + \makebox[0.5cm]{3.2m}}} & \small{\makebox[0.7cm]{0.213} \, \textcolor{gray}{\makebox[0.5cm]{1.1m} + \makebox[0.5cm]{3.2m}}} \\
  \small{\CornellBox{}} & \small{\makebox[0.7cm]{0.032} \, \textcolor{gray}{\makebox[0.5cm]{31s} + \makebox[0.5cm]{1.1m}}} & \small{\makebox[0.7cm]{\textbf{0.029}} \, \textcolor{gray}{\makebox[0.5cm]{31s} + \makebox[0.5cm]{1.1m}}} \\
  \small{\GlossyKitchen{}} & \small{\makebox[0.7cm]{\textbf{0.176}} \, \textcolor{gray}{\makebox[0.5cm]{1.1m} + \makebox[0.5cm]{3.3m}}} & \small{\makebox[0.7cm]{0.224} \, \textcolor{gray}{\makebox[0.5cm]{1.1m} + \makebox[0.5cm]{3.3m}}} \\
  \small{\Necklace{}} & \small{\makebox[0.7cm]{\textbf{0.115}} \, \textcolor{gray}{\makebox[0.5cm]{27s} + \makebox[0.5cm]{1.0m}}} & \small{\makebox[0.7cm]{0.183} \, \textcolor{gray}{\makebox[0.5cm]{27s} + \makebox[0.5cm]{1.0m}}} \\
  \small{\SwimmingPool{}} & \small{\makebox[0.7cm]{\textbf{0.073}} \, \textcolor{gray}{\makebox[0.5cm]{55s} + \makebox[0.5cm]{2.0m}}} & \small{\makebox[0.7cm]{0.073} \, \textcolor{gray}{\makebox[0.5cm]{55s} + \makebox[0.5cm]{2.0m}}} \\
  \small{\Torus{}} & \small{\makebox[0.7cm]{0.078} \, \textcolor{gray}{\makebox[0.5cm]{34s} + \makebox[0.5cm]{59s}}} & \small{\makebox[0.7cm]{\textbf{0.074}} \, \textcolor{gray}{\makebox[0.5cm]{34s} + \makebox[0.5cm]{59s}}} \\
  \small{\VeachDoor{}} & \small{\makebox[0.7cm]{\textbf{0.182}} \, \textcolor{gray}{\makebox[0.5cm]{1.0m} + \makebox[0.5cm]{2.6m}}} & \small{\makebox[0.7cm]{0.206} \, \textcolor{gray}{\makebox[0.5cm]{1.0m} + \makebox[0.5cm]{2.6m}}} \\
  \small{\WaterCaustic{}} & \small{\makebox[0.7cm]{0.515} \, \textcolor{gray}{\makebox[0.5cm]{37s} + \makebox[0.5cm]{2.0m}}} & \small{\makebox[0.7cm]{\textbf{0.395}} \, \textcolor{gray}{\makebox[0.5cm]{37s} + \makebox[0.5cm]{2.0m}}} \\
  \small{\GlossyCornellBox{}} & \small{\makebox[0.7cm]{\textbf{0.056}} \, \textcolor{gray}{\makebox[0.5cm]{36s} + \makebox[0.5cm]{1.8m}}} & \small{\makebox[0.7cm]{0.070} \, \textcolor{gray}{\makebox[0.5cm]{36s} + \makebox[0.5cm]{1.8m}}} \\
  \bottomrule
\end{tabular*}

    \vspace{0mm}
\end{minipage}
    \setlength{\fboxrule}{10pt}%
\setlength{\insetvsep}{20pt}%
\setlength{\tabcolsep}{-1pt}%
\renewcommand{\arraystretch}{1}%
\small%
\begin{tabular}{cccc}
  & Ruppert et al. & SDMM (Ours) & Reference \\
    \setInset{A}{red}{45}{143}{60}{60}%
    \setInset{B}{orange}{66}{160}{60}{60}%
    \addBeautyCrop{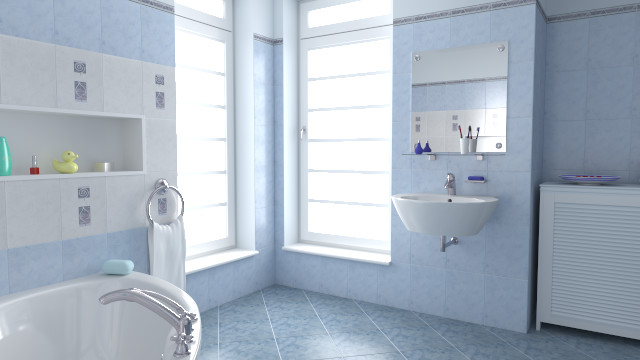}{0.19}{640}{360}{0}{7}{346}{344} &
    \addInsets{images/fig-vmm/glossy-bathroom-VMM_u.jpg} &
    \addInsets{images/fig-vmm/glossy-bathroom-SDMM-radiance_u.jpg} &
    \addInsets{images/fig-references/glossy-bathroom-reference.jpg} \\
  \multicolumn{1}{r}{MAPE:} & \textbf{0.104} & 0.179 & \\
  \multicolumn{1}{r}{training + rendering time:} & 1.8m + 4.7m & 1.8m + 4.7m & \\
  \multicolumn{1}{r}{training + rendering spp:} & 508 + 1480 & 252 + 772 & \\
    \setInset{A}{red}{240}{13}{60}{60}%
    \setInset{B}{orange}{383}{243}{60}{60}%
    \addBeautyCrop{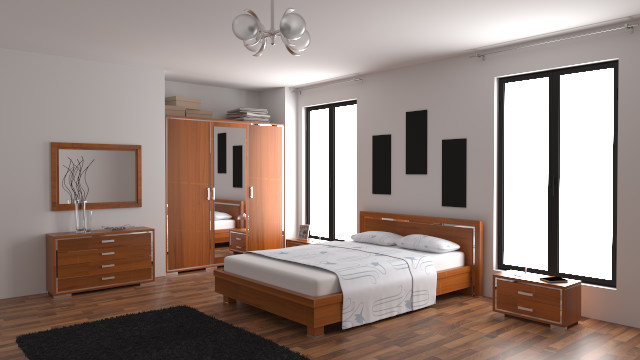}{0.19}{640}{360}{147}{7}{346}{344} &
    \addInsets{images/fig-vmm/bedroom-VMM_u.jpg} &
    \addInsets{images/fig-vmm/bedroom-SDMM-radiance_u.jpg} &
    \addInsets{images/fig-references/bedroom-reference.jpg} \\
  \multicolumn{1}{r}{MAPE:} & \textbf{0.054} & 0.059 & \\
  \multicolumn{1}{r}{training + rendering time:} & 1.4m + 3.0m & 1.4m + 3.0m & \\
  \multicolumn{1}{r}{training + rendering spp:} & 428 + 1112 & 252 + 772 & \\
    \setInset{A}{red}{300}{100}{60}{60}%
    \setInset{B}{orange}{350}{283}{60}{60}%
    \addBeautyCrop{images/fig-references/water-caustic-reference-lq.jpg}{0.19}{640}{360}{147}{7}{346}{344} &
    \addInsets{images/fig-vmm/water-caustic-VMM_u.jpg} &
    \addInsets{images/fig-vmm/water-caustic-SDMM-radiance_u.jpg} &
    \addInsets{images/fig-references/water-caustic-reference.jpg} \\
  \multicolumn{1}{r}{MAPE:} & 0.515 & \textbf{0.395} & \\
  \multicolumn{1}{r}{training + rendering time:} & 37s + 2.0m & 37s + 2.0m & \\
  \multicolumn{1}{r}{training + rendering spp:} & 436 + 1592 & 252 + 772 & \\
\end{tabular}

    \captionof{figure}{\label{fig:vmm}
        Visual comparison of the same experimental setup as in \autoreftable{tab:vmm} for three selected scenes.
    }
\end{figure}

\ADD{We compare our method with the improved version \cite{mueller19guiding} of ``Practical Path Guiding'' (PPG)~\cite{mueller2017practical} as well as with parallax-aware von Mises-Fisher mixtures~\cite{ruppert2020robust} trained on radiance (VMM radiance).

When comparing with PPG, we disable PPG's learning of the BSDF selection probability to ensure a fair comparison to our work, which does not currently have this feature, but could be extended with it.
Instead, we use the fixed BSDF sampling fraction of $50\%$ for PPG and for our radiance-based guiding scheme (SDMM radiance), and $30\%$ for our product-based guiding scheme (SDMM product).

Like PPG, our method automatically chooses when to terminate training based on the overall rendering budget and is designed to include all iterations---training and rendering---in the final image.
The implementation of Ruppert et al.~\cite{ruppert2020robust}, however, has distinct training and rendering components, which puts it at a disadvantage in online-training comparisons.
Therefore, in addition to comparing with Ruppert et al.~\cite{ruppert2020robust} at equal sample counts (\autoreftable{tab:results} and \autoref{fig:images}) in an online-training setting, we provide an equal training- and rendering time comparison in \autoreftable{tab:vmm} and \autoref{fig:vmm}.}

\paragraph*{Online training comparisons.}
In \autoreftable{tab:results} and \autoref{fig:images} we compare PPG, radiance-based VMMs, and our method with (SDMM product) and without (SDMM radiance) product guiding, on $12$ virtual scenes that have varied illumination characteristics.
For each method and scene, we report the mean absolute percentage error (MAPE), the render time, and the speedup vs.\ PPG at reaching the lowest common MAPE value.
MAPE is defined as $\frac{1}{N} \sum_{i=1}^N |v_i - \hat{v}_i| / (\hat{v}_i + \epsilon)$, where $\hat{v}_i$ is the value of the $i$-th pixel in the reference image, $v_i$ is the value of the $i$-th rendered pixel, and $\epsilon = 0.01$ prevents near-black pixels from dominating the metric.
A $2\times$ smaller MAPE loosely corresponds to $4\times$ faster rendering.

A common trend of our approach is that it achieves the lower per-sample error than PPG in all scenes and achieves slightly lower error than radiance-based VMMs in $8$ our of $12$ scenes.
Product-based guiding further reduces the error, but at great computational cost.

As expected, our method performs well when there exists a large degree of spatio-directional correlation in the incident radiance, e.g.\ as induced by small, local luminaires in the \Bookshelf{}, the \CornellBox{}, the \WaterCaustic{}, and the \GlossyCornellBox{} scenes; see the insets in \autoref{fig:images}.

However, the computational overhead of our technique is significantly larger than that of PPG and VMMs.\@
Taking the render time into account and comparing \emph{time to equal error}, our radiance- and product-based guiding approaches only outperform PPG in $6$ out of $12$ scenes, with our radiance- and product-based guiding performing competitively with each other.
In \autoref{fig:convergence}, we reinforce this observation by plotting MAPE against render time and against the number of samples per pixel.
We report additional metrics and false-color error visualizations in our supplementary HTML-based results viewer.

\begin{figure*}
    \setlength{\fboxrule}{1pt}%
\setlength{\insetvsep}{20pt}%
\setlength{\tabcolsep}{1pt}%
\renewcommand{\arraystretch}{0.75}%
\small%
\hspace*{-0.5mm}%
\begin{tabular}{ccccccccc}
  \multicolumn{3}{c}{\Bedroom{}}&\multicolumn{3}{c}{\Bookshelf{}}&\multicolumn{3}{c}{\GlossyCornellBox{}}\\
  \multicolumn{3}{c}{\begin{overpic}[width=0.33\textwidth]{images/fig-references/bedroom-reference-lq.jpg}
\put(77.76041666666667,2.994791666666667){\makebox(0,0){\tikz\draw[red,fill=red] (0,0) circle (0.0033\textwidth);}}
\put(27.500000000000004,50.208333333333336){\makebox(0,0){\tikz\draw[orange,fill=orange] (0,0) circle (0.0033\textwidth);}}
\end{overpic}}&\multicolumn{3}{c}{\begin{overpic}[width=0.33\textwidth]{images/fig-references/bookshelf-reference-lq.jpg}
\put(32.8125,44.114583333333336){\makebox(0,0){\tikz\draw[red,fill=red] (0,0) circle (0.0033\textwidth);}}
\put(73.69791666666666,20.520833333333332){\makebox(0,0){\tikz\draw[orange,fill=orange] (0,0) circle (0.0033\textwidth);}}
\end{overpic}}&\multicolumn{3}{c}{\begin{overpic}[width=0.33\textwidth]{images/fig-references/glossy-cbox-reference-lq.jpg}
\put(60.55572916666667,44.097239583333334){\makebox(0,0){\tikz\draw[red,fill=red] (0,0) circle (0.0033\textwidth);}}
\put(56.944270833333334,18.61109375){\makebox(0,0){\tikz\draw[orange,fill=orange] (0,0) circle (0.0033\textwidth);}}
\end{overpic}}\\
  \color{red}\fbox{\includegraphics[width=0.10333333333333333\textwidth]{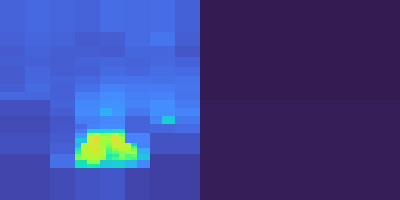}}&\color{red}\fbox{\includegraphics[width=0.10333333333333333\textwidth]{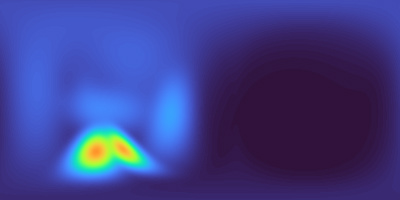}}&\color{red}\fbox{\includegraphics[width=0.10333333333333333\textwidth]{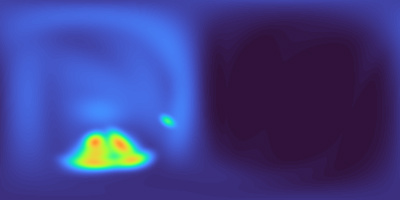}}&\color{red}\fbox{\includegraphics[width=0.10333333333333333\textwidth]{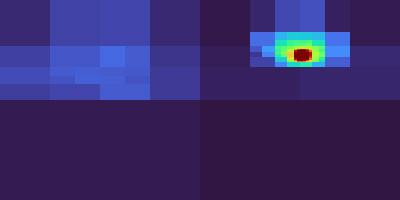}}&\color{red}\fbox{\includegraphics[width=0.10333333333333333\textwidth]{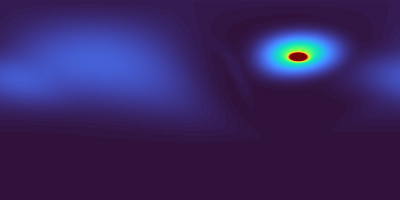}}&\color{red}\fbox{\includegraphics[width=0.10333333333333333\textwidth]{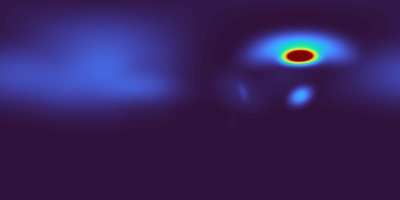}}&\color{red}\fbox{\includegraphics[width=0.10333333333333333\textwidth]{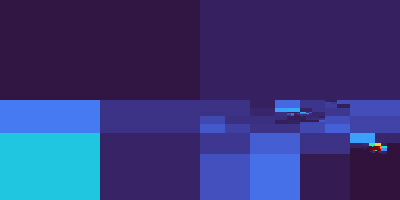}}&\color{red}\fbox{\includegraphics[width=0.10333333333333333\textwidth]{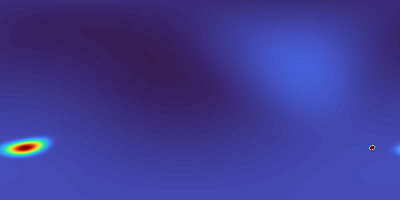}}&\color{red}\fbox{\includegraphics[width=0.10333333333333333\textwidth]{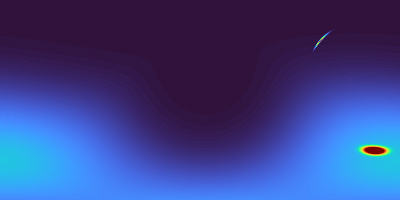}}\\
  \color{orange}\fbox{\includegraphics[width=0.10333333333333333\textwidth]{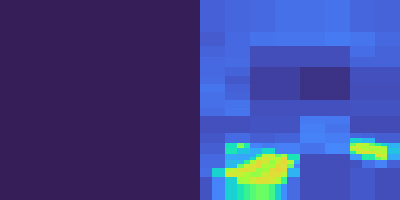}}&\color{orange}\fbox{\includegraphics[width=0.10333333333333333\textwidth]{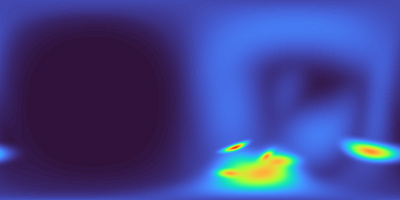}}&\color{orange}\fbox{\includegraphics[width=0.10333333333333333\textwidth]{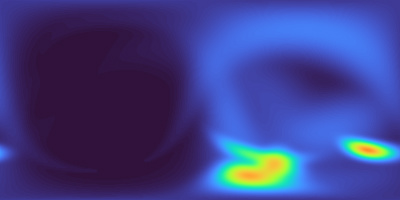}}&\color{orange}\fbox{\includegraphics[width=0.10333333333333333\textwidth]{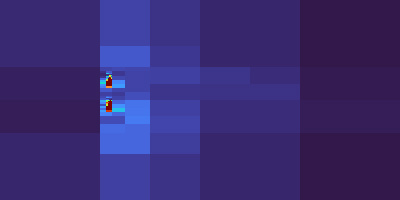}}&\color{orange}\fbox{\includegraphics[width=0.10333333333333333\textwidth]{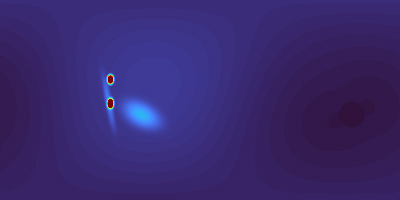}}&\color{orange}\fbox{\includegraphics[width=0.10333333333333333\textwidth]{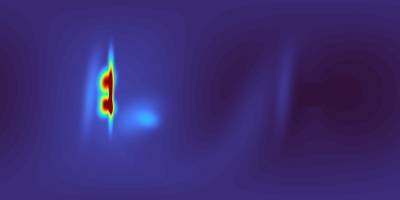}}&\color{orange}\fbox{\includegraphics[width=0.10333333333333333\textwidth]{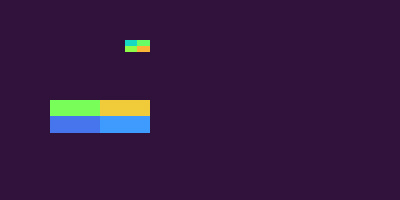}}&\color{orange}\fbox{\includegraphics[width=0.10333333333333333\textwidth]{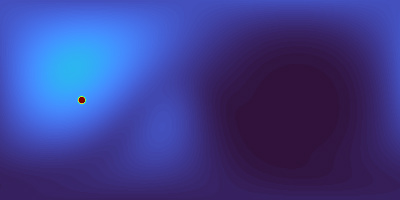}}&\color{orange}\fbox{\includegraphics[width=0.10333333333333333\textwidth]{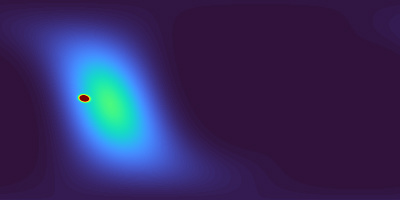}}\\
  PPG&SDMM&DMM&PPG&SDMM&DMM&PPG&SDMM&DMM\\
\end{tabular}

    \vspace{-1mm}
    \caption{\label{fig:visualizer}
        Learned directional PDFs, conditioned on two spatial locations per scene, visualized as false-color images in \ADD{spherical} coordinates.
        We compare our learned distributions (SDMM) with those learned by PPG and by a purely directional version of our mixture model (DMM).
        As expected, the SDMMs produce sharper distributions than the spatially marginalized DMMs without exhibiting discretization artifacts like PPG does---especially in the difficult \GlossyCornellBox{} scene.
        Zooming in is recommended.
        The SDMMs also smoothly capture high-frequency spatio-directional correlation, which we illustrate in our supplementary video by smoothly varying the queried spatial coordinate.
    }
\end{figure*}

\paragraph*{Offline training comparisons.}
\ADD{To provide a fair equal-time ecomparison with Ruppert et al.~\cite{ruppert2020robust}, we match their experimental setup where each method trains and renders for a fixed, equal time; see \autoreftable{tab:vmm} and \autoref{fig:vmm}.
We have made no attempt to fine-tune our method to this paradigm, and no attempt to modify their method to include training iterations into the final image.
Instead, we use both methods as-is, and compare the images generated after training has been terminated in both methods.
To ensure fairness, we disabled the inverse-variance based combination of iterations for our method and instead weigh all samples equally.

We note that Ruppert et al.~\cite{ruppert2020robust} outperform our algorithm on the majority of the scenes.
Their algorithm is significantly faster than ours, meaning that it is able to render far more samples in the same time compared as ours.
Thus, even though the quality of learned mixture models is comparable---as evidenced by the equal sample count comparison \autoref{fig:images}---their method is more practical in most of our test scenes.}

\paragraph*{Quality of the guiding distribution.}
In \autoref{fig:visualizer}, we visualize the learned spatio-directional mixture model in $3$ scenes.
For each scene, we show the $2$D distributions obtained by conditioning the mixture model on $2$ indicated spatial locations.
The mixture model is not only accurate, but it also captures high-frequency spatio-directional correlation, which we illustrate in our supplementary video by smoothly varying the spatial coordinate that the model is conditioned on.

\section{Discussion and Future Work}%
\label{sec:discussion}

\paragraph*{Mini-batch EM versus stepwise EM.}
Stepwise EM~\cite{Capp__2009}, later extended to robustly handle weighted samples by Vorba et al.~\cite{Vorba:2014:OnlineLearningPMMinLTS}, is an alternative online EM algorithm that could be used instead of mini-batch EM to train SDMM.\@
The difference between the two algorithms is the frequency of the Robbins-Monroe update of the sufficient statistics \autoref{eq:robbins-monroe}:
in stepwise EM, the update is performed per sample, whereas in mini-batch EM, it is performed per mini-batch.

Because the aforementioned update is expensive in tangent spaces (due to \autoref{eq:robbins-monroe-2nd}), we prefer using mini-batch EM.\@
Another argument in favor of mini-batch EM is that stepwise EM unduly weights earlier samples within the same batch higher due to them experiencing the Robbins-Monroe update earlier, despite them being sampled from the same distribution.
This uneven averaging of the sufficient statistics of the batch unnecessarily increases the variance of the optimization.

\paragraph*{Practicality of SDMM.}
Compared with PPG~\cite{mueller19guiding,mueller2017practical}, we demonstrated superior equal-sample-count error of both our radiance- and product-based guiding approaches.
However, the computational overhead of our Mitsuba implementation results in better overall efficiency on only a subset of the scenes.
\ADD{Compared with VMMs~\cite{ruppert2020robust}, the difference is even larger: while we achieve similar error at equal sample counts, their lower computational cost leads to better efficiency than ours in most scenes.
The practicality of spatio-directional mixture models is thus limited and further research into efficient implementations and approximations is needed.}

To this end, we believe there are a promising number of optimizations that can still be made, such as automatic pruning of the product mixture~\cite{herholz2016}.
Finally, the computational bottleneck may shift when more complex scenes are rendered, since the cost of ray tracing and shading will be larger.

\begin{figure}
    \setlength{\fboxrule}{10pt}%
    \setlength{\insetvsep}{20pt}%
    \setlength{\tabcolsep}{1.8pt}%
    \renewcommand{\arraystretch}{1}%
    \small%
    \begin{tabular}{cc}
        VMM radiance~\cite{ruppert2020robust} &
      SDMM radiance (Ours) \\
      \includegraphics[width=0.49\columnwidth,trim=0 90 75 150,clip]{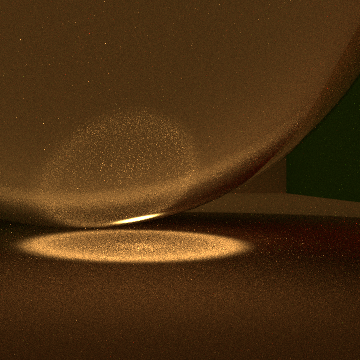} &
      \includegraphics[width=0.49\columnwidth,trim=0 90 75 150,clip]{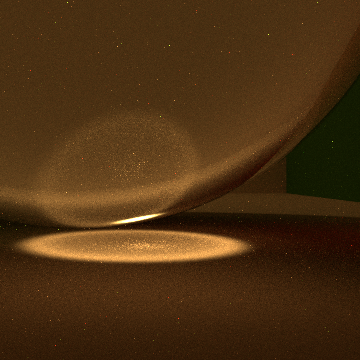} \\
      \includegraphics[width=0.49\columnwidth,trim=0 90 75 150,clip]{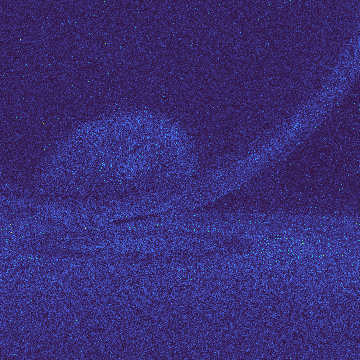} &
      \includegraphics[width=0.49\columnwidth,trim=0 90 75 150,clip]{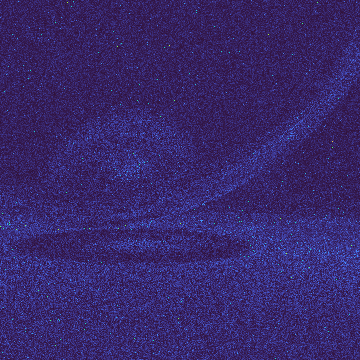} \\
      MAPE: $0.172$ &
      MAPE: $\mathbf{0.130}$ \\
    \end{tabular}
    \caption{\label{fig:zoomed}
        \ADD{While the parallax heuristic of Ruppert et al.~\cite{ruppert2020robust} is cheap and accurate in most situations, certain effects, such as lensing, can not only result in non-linear parallax, but even invert its direction.
        Such effects are rarely relevant in practice, but we nonetheless show one such occurrence: the caustic in the \GlossyCornellBox{} scene.
        The lower noise of our method at an equal number of samples is indicative of a locally more accurate fit to incident radiance.}
    }
\end{figure}

\paragraph*{Combination with Reprojection.}
\ADD{
While our data-driven approach of learning $5$D mixtures may seem in contrast with the reprojection of $2$D mixtures~\cite{ruppert2020robust}, we believe that the two approaches are not mutually exclusive.
In particular, there are situations in which the reprojection heuristic breaks down.
Lensing effects are one such example, in which the appropriate hemispherical movement is actually the \emph{reverse} of what reprojection would predict; we show an example in \autoref{fig:zoomed}.
While lensing effects occur relatively rarely in practice, this example demonstrates one the benefits of using a data-driven approach compared to hand-crafted hauristics.
Thus, we believe that one could use $5$D mixtures to learn an \emph{offset} on top of the reprojections of Ruppert et al.~\cite{ruppert2020robust}, fixing the rare special cases where the assumptions of the reprojection are incorrect.
}

\paragraph*{Computational implications of tangent spaces.}
Although product sampling is enabled by parameterizing each Gaussian in its own tangent space, such a per-Gaussian parameterization necessitates the on-the-fly computation of numerous changes of variables:
a change of variables is needed whenever multiple Gaussians interact, regardless of whether this interaction amounts to conditioning, product sampling, or EM optimization.
Despite the optimizations that we already perform, these changes of variable are the computational bottleneck of our implementation.
Thus, the efficiency of our method could be greatly improved by coming up with means to reduce the number of changes of variables, e.g.\ through a different mixture model, or to compute them more efficiently.

\paragraph*{Alternative acceleration structures}
\ADD{
The choice of placing our Gaussians within a $k$D-tree was motivated by the intractable quadratic cost of using a single global mixture; see also \autoref{fig:mixture-comparison}.
Alternatively, the quadratic cost could also be mitigated by ignoring Gaussians whose density is below a threshold, e.g.\ by placing the Gaussians in a BVH, where the bounding volumes contain a large percentage of each Gaussians' mass.
However, such an approach maintains two disadvantages over our $k$D-tree approach:
first, using a global mixture may lead to Gaussians that cover significant portions of the scene, which result in quadratic cost regardless of acceleration structure.
And second, thresholding the size of Gaussians without introducing large error is non-trivial, because the conditioning step may amplify small densities unpredictably.
We implemented a prototype and experimentally confirmed these difficulties \cite{Dodik2020PathGU}.}

\paragraph*{Extensions.}
Our implementation does not perform a number of well-established extensions to path-guiding techniques to facilitate a simple comparison with previous work.
When using our algorithm in practice, such orthogonal extensions should, however, be used.
First, next-event estimation (NEE) should be performed and the learned guiding distribution should be optimized to \emph{complement} NEE~\cite{mueller2019nis}.
In the context of Gaussian mixtures, Ruppert et al.~\cite{ruppert2020robust} describe an appropriate modification to the Monte Carlo samples that the model is trained on.
Second, the BSDF sampling fraction should be learned~\cite{mueller19guiding} as opposed to set to a fixed constant.
Third, adjoint-driven Russian roulette and splitting~\cite{VorbaKrivanek:2016:ADRRS} should be performed.

Furthermore, there are also mixture-specific techniques that could conceivably enhance our algorithm.
For example, careful splitting and merging of mixture components could simultaneously increase our model capacity as well as escape local minima in the EM optimization~\cite{ruppert2020robust}.
We suspect that the splitting might only be necessary along the directional dimensions of the SDMMs, since the kD-tree data-structure already performs spatial splitting.

\paragraph*{High-dimensional mixtures.}
We demonstrated the ability of Gaussian mixtures to directly approximate the $5$D incident radiance as well as certain $n$D BSDF models.
However, the BSDF models that we used were relatively simple, never exceeding ${n = 5}$.
In the future, we would like to investigate the use of higher-dimensional mixtures to capture additional variation, such as found in the $10$D Disney BSDF~\cite{Burley2012PhysicallyBasedSA}, or additional dimensions in the incident radiance, such as the wavelength in spectral rendering.
Such higher-dimensional mixtures need advanced data structures and a form of sparsity to combat the curse of dimensionality---a good starting point is likely the manifold framework of Herholz et al.~\cite{Herholz:2018}.

\paragraph*{Non-linear spatio-directional dependencies.}
Our spatio-directional Gaussians explicitly model correlations (i.e.\ linear relationships) between the $5$ dimensions of the incident radiance field.
One example of such a linear relationship if parallax.
However, the spatio-directional dimensions often also depend on each other \emph{non-linearly}, which can not be modeled by Gaussian covariance matrices.
A mixture of more sophisticated distributions may thus exhibit more accurate fits with fewer mixture components.

\section{Conclusion}

We demonstrated the feasibility of using spatio-directional mixture models for path guiding.
In particular, we showed that a tangent-space parameterization of Gaussians enables product sampling among a $5$D mixture that approximates incident radiance and $n$D mixtures that approximate the BSDFs.
The use of thousands of mixture components---a necessity for accurately modeling the intricate radiance field---was made practical by a $k$D-tree data structure.

\ADD{The tangent-space Gaussian mixture performed remarkably well in our experiments, making us hopeful that it can become an alternative to von Mises-Fisher mixtures, featuring anisotropy in addition to rotational symmetry.}

\ADD{As for data-driven $5$D spatio-directional mixtures: they achieve competetive results to previous work in equal sample count comparisons, but are outperformed by other spatio-directional models that recover spatio-directional correlation from reprojection~\cite{ruppert2020robust}.
Nonetheless, data-driven SDMMs help in the cases where reprojection is inaccurate, such as lensing effects by curved, specular geometries.
This makes us hopeful that---if the overhead can be further reduced---SDMMs can become a practical tool on a practitioner's toolbelt, e.g.\ by combining them with reprojection.}

\section*{Acknowledgments}
We thank Marco Manzi for creating \autoref{fig:2d-indicend-radiance} and Alex Keller for valuable feedback.
We also thank the following people for providing scenes that appear in our figures:
Benedikt Bitterli~\cite{resources16},
Johannes Hanika (\Necklace),
Miika Aittala, Samuli Laine, and Jaakko Lehtinen (\VeachDoor),
Olesya Jakob (\Torus),
Ond\v{r}ej Karl\'{\i}k (\SwimmingPool),
SlykDrako (\Bedroom), and
Tiziano Portenier (\Bathroom, \Bookshelf).
Lastly, we thank Markus Gross for enabling and supporting this research.

This work was supported by a UKRI Future Leaders Fellowship [grant number G104084].


\printbibliography


\appendix

\section{Jacobian Matrices}
\label{app:jacobians}

The Jacobian matrices of the $\mu$-centered $\toTangent_\mu$ and $\toWorld_\mu$ maps can be derived by differentiating Equations~\eqref{eq:log} and \eqref{eq:exp}, resulting in
\begin{align}
    J_{\toTangent_\mu}(\dir) &= &&\begin{bmatrix}
        \mathcal{J}_{\toTangent_\mu} & 0 & \dir^{\circlearrowleft}_x \frac{\dir^{\circlearrowleft}_z - \mathcal{J}_{\toTangent_\mu}}{(1 - \dir^{\circlearrowleft}_z\dir^{\circlearrowleft}_z) \mathcal{J}_{\toTangent_\mu}} \\
        0 & \mathcal{J}_{\toTangent_\mu} & \dir^{\circlearrowleft}_y \frac{\dir^{\circlearrowleft}_z - \mathcal{J}_{\toTangent_\mu}}{(1 - \dir^{\circlearrowleft}_z\dir^{\circlearrowleft}_z) \mathcal{J}_{\toTangent_\mu}}
    \end{bmatrix} R_\mu \, \\
    J_{\toWorld_\mu}(\nu) &= R_\mu^{-1} &&\begin{bmatrix}
        \sinc{\|\nu\|} + \nu_x^2 \mathcal{J}_{\toWorld_\mu} & \sinc{\|\nu\|} + \nu_x\nu_y \mathcal{J}_{\toWorld_\mu} \\
        \sinc{\|\nu\|} + \nu_y\nu_x \mathcal{J}_{\toWorld_\mu} & \sinc{\|\nu\|} + \nu_y \mathcal{J}_{\toWorld_\mu} \\
        -\nu_x \sinc{\|\nu\|} & -\nu_y \sinc{\|\nu\|}
    \end{bmatrix} \,,
\end{align}
\begin{align}
    \mathcal{J}_{\toTangent_\mu} = \frac{1}{\sinc(\cos^{-1}(\dir^{\circlearrowleft}_z))} \,, \,\, \mathcal{J}_{\toWorld_\mu} = \frac{\cos{\|\nu\|} - \sinc{\|\nu\|}}{\|\nu\|^2} \,,
\end{align}
where $\dir^{\circlearrowleft} = R_\mu \dir$ and $R_\mu$ is the rotation matrix that rotates $\mu$ to ${(0, 0, 1)}$.
We use the \emph{unnormalized} $\sinc$ function.

\section{Covariance Initialization}
\label{app:cov-init}

Given an initial $5$D mean-vector $(\pos,\dir)$ of a Gaussian mixture component, we wish to initialize its covariance matrix such that it is
\begin{itemize}
    \item isotropic in the tangent plane of the surface at $\pos$ and in the tangent space around $\dir$, and
    \item flat along the surface normal at $\pos$.
\end{itemize}
To this end, we let $90\%$ of the mass of each Gaussian to be contained within a certain radius, $r_{st}$ in the surface's tangent plane, and $r_{n}$ along the normal.
More concretely, we compute the initial spatial ${3\times3}$ part of the covariance matrix as
\begin{align}
    \Sigma^\pos = \frac{r_{st}^2 \mathbf{s} \mathbf{s}^T + r_{st}^2 \mathbf{t} \mathbf{t}^T + r_{n}^2 \mathbf{n} \mathbf{n}^T}{{\chi^2_3}^{-1}(0.9)},
\end{align}
where $\mathbf{s}, \mathbf{t}, \mathbf{n}$ form the basis of the local coordinate frame at the selected point and ${\chi^2_3}^{-1}$ is the inverse cumulative distribution function of the chi-square distribution.
We empirically found good results with the values $r_n = 3 \cdot 10^{-2}$, and $r_{st} = 2 \cdot \frac{d}{2}$, where $d$ is the length of the longest side of the leaf node containing the Gaussian and the division by $2$ is due to the number of distinct spatial mean vectors in the mixture.

The initial directional ${2\times2}$ part $\Sigma^\dir$ of the covariance matrix, living in the $\dir$-centered tangent space, is set to be isotropic and inversely proportional to the number of directional Gaussian components at a spacial location (in our case $8$), with the diagonal entries being equal to $\sigma^2 = \frac{2\pi}{8}$, resulting in the full covariance matrix
\begin{align}
    \Sigma = \begin{bmatrix}
        \Sigma^\dir & \mathbf{0} \\
        \mathbf{0} & \Sigma^\pos
    \end{bmatrix} \,.
\end{align}

\section{Modified \texttt{k-means++} Algorithm}%
\label{app:k-means}

The modification to the original \texttt{k-means++} algorithm consists of defining a custom distance metric $D(\pos_i,\pos_j)$ between spatial positions $\pos_i$ and $\pos_j$ that takes their surface normals $\normal_i$ and $\normal_j$ into account, as well as defining a custom resampling probability $p^{++}(\pos_i)$ that takes the corresponding Monte Carlo weight $w_i$ of the sample at $\pos_i$ into account.

Our custom distance metric extends the spatial euclidean distance by the geodesic distance between normals:
\begin{align}
    \text{dist}(\pos_i, \pos_j)^2 &= \left( \cos^{-1}(\normal_i \cdot \normal_j) \,/\, \pi \right)^2 + \| \pos_i - \pos_j \|^2 \,,
\end{align}
where the division by $\pi$ normalizes geodesic distances to the range $[0, 1]$.
Additionally, we enforce a poisson-disk distribution of points by setting the distance to zero within pre-defined spatial and spherical radii $T_\pos$ and $T_\normal$:
\begin{align}
    D(\pos_i, \pos_j) &= \begin{cases}
        \text{dist}(\pos_i, \pos_j) & \mbox{if } \| \pos_i - \pos_j \| > T_\normal \vee \cos^{-1}(\normal_i \cdot \normal_j) > T_\pos \,, \\
        0 & \mbox{otherwise} \,.
    \end{cases}
\end{align}
We empirically choose ${T_\normal = 0.2}$ and ${T_\pos = 1.6 \times 10^{-3}}$.

Building on top of the custom distance metric, we extend the resampling probability $p^{++}(\pos_i)$ with the Monte Carlo weight $w_i$ by multiplying the distance to the closest neighbor $\pos_j$ by the clamped Monte Carlo weight in the interval $[W_0, W_1]$:
\begin{align}
    p^{++}(\pos_i) = \text{clamp}(w_i, W_0, W_1) \min_{j \neq i} D(\pos_i, \pos_j) \,.
\end{align}
The clamping to the interval $[W_0, W_1]$ serves the double-purpose of allowing zero-valued (black) samples to contribute when fewer than 3 radiance-carrying samples are nearby, as well as preventing fireflies from dominating other samples.
We empirically found ${W_0 = 10^{-1}}$ and ${W_1 = 3}$ to work well.

\end{document}